\documentclass[12pt]{article} 
\pdfoutput=1
\usepackage{amssymb,graphicx}
\usepackage{epstopdf}
\usepackage{amsmath,amsfonts}
\usepackage{epsfig} 
\usepackage{graphicx,graphics}
\usepackage[dvipsnames]{xcolor}

\usepackage[unicode=true,pdfusetitle,
bookmarks=true,bookmarksnumbered=true,bookmarksopen=true,bookmarksopenlevel=3,
 breaklinks=false,pdfborder={0 0 1},backref=false,colorlinks=true,citecolor=blue]
 {hyperref}

\begin{document}

\title{\bf Domain walls through different cosmologies}
\author{I.~Dankovsky$^{a,b}$, D.~Gorbunov$^{b,c}$, S.~Ramazanov$^d$, A.~Vikman$^e$\\
\small{\em $^a$Faculty of Physics, Lomonosov Moscow State University, 119991 Moscow, Russia}\\
\small{\em $^b$Institute for Nuclear Research of the Russian Academy of Sciences, 117312 Moscow, Russia}\\
\small{\em  $^c$Moscow Institute of Physics and Technology, 141700 Dolgoprudny, Russia}\\ 
\small{\em  $^d$Institute for Theoretical and Mathematical Physics, MSU, 119991 Moscow, Russia}\\ 
\small{\em  $^e$CEICO, FZU-Institute of Physics of Czech Academy of Sciences,}\\
\small{\em Na Slovance 1999/2, 182 00 Prague, Czech Republic}}

 \date{}

{\let\newpage\relax\maketitle}

\begin{abstract}
We study properties of domain walls (DWs) arising in the model with a double well potential assuming different early universe cosmologies: from dust through stiff matter domination to the limit of effective Minkowski space. Using lattice simulations we demonstrate that evolution of DW networks exhibits an approximate universality with respect to the cosmological equation of state (EoS). 
Namely, the wall area inside a given large volume is mainly determined by the particle horizon, while details of cosmic expansion play a subdominant role. As it follows, particle horizon rather than the inverse Hubble rate is pivotal in DW evolution defining the network correlation length and hence its phenomenology, e.g., the characteristic wavelength of emitted gravitational waves (GWs). Formation of closed DWs is shown to be very sensitive to the cosmological EoS and hence violate the universality, but their contribution to the total network is too small to change the overall picture. The universality breaking is also observed in the spectral properties of GWs from annihilating (biased) DWs: i) the IR slope of the spectrum depends on the EoS parameter; ii) there is a plateau in the UV part of the spectrum, which gets more pronounced as one stiffens the EoS. However, the dependence on the EoS is rather weak in the near peak region, which is most relevant from the viewpoint of pulsar timing arrays and other searches for the stochastic GW background. For a selection of primordial cosmologies, we provide fitting formulae for the spectra near the peak.

\end{abstract}

%%%%%%%%%%%%%%%%%%%%%%%%%%%%%%%%%%%%%%%%%%%%%%%%%%%%%%%%%%%%%%%%%%%%%%%%%%%%
\section{Introduction} 
\label{sec:intro}

Cosmic domain walls (DWs) are generally predicted in particle physics models exhibiting spontaneous breaking of discrete symmetries~\cite{Zeldovich:1974uw, Kibble:1976sj}, and therefore they are worth a detailed analysis. Most commonly, it is assumed that DWs spend their entire life, from birth at a phase transition till annihilation, during radiation domination (RD). Many properties of DW evolution and the resulting gravitational wave (GW) emission have been uncovered with the use of lattice simulations in this setup. In particular, it has been established that the DW network settles to the self-similar evolution some time after its formation~\cite{Press:1989yh, Garagounis:2002kt, Hiramatsu:2013qaa}. In this scaling regime, the network is dominated by a single long wall with a curvature radius being of the order of the cosmic time. There are several reasons to extend the study of DWs to cosmologies, where the dominant component described by the pressure ${\cal P}_{U}$ and energy density $\rho_{U}$ exhibits other than the radiation-like equation of state (EoS), ${\cal P}_{U}=w\rho_{U}$, $w \neq 1/3$. First, little is known about the universe EoS prior to Big Bang Nucleosynthesis (BBN). Second, existence of the primordial stage with the EoS parameter $w>1/3$ can significantly amplify the GW signal produced by DWs. Then, it is important to test the level of universality/diversity of DW phenomenology with respect to different possible early universe histories. Last but not least, studying, how evolution of the DW network reflects variations in the parameter $w$ can also help elucidate theoretical aspects of DW physics. 

We assume the simplest DW model involving a double well potential. We study how the DW network  evolves in the spatially flat FLRW universe filled with the barotropic fluid described by a constant EoS parameter $w$, whose energy density decreases as 
\begin{equation}
\label{1st}
\rho_U \propto \frac{1}{a^{3(1+w)}} \; ,
\end{equation}
where $a$ is the universe expansion factor. For this purpose we carry out numerical simulations using CosmoLattice code~\cite{Figueroa:2020rrl, Figueroa:2021yhd, Baeza-Ballesteros:2025tme}; the similar numerical analysis has been recently undertaken in Ref.~\cite{Blasi:2025tmn}. Earlier aspects of DW evolution in non-RD like backgrounds have been studied in Refs.~\cite{Garagounis:2002kt, Avelino:2005kn, Leite:2011sc, Leite:2012vn, Martins:2016ois}, which employ the Press-Ryden-Spergel (PRS) approach~\cite{Press:1989yh} allowing for an extended time run of simulations by the price of modifying the equation of motion for the DW field. We do not follow this approach and do not alter the equation of motion for the relevant scalar. It is also worth mentioning
Ref.~\cite{Hindmarsh:1996xv} attempting at the analytical description of DWs for different values of the EoS parameter. Whenever relevant, we compare our results with the results obtained in these earlier works.

Crucially, we observe universality of DW scaling with respect to background cosmology. In particular, the DW comoving area $S$ encapsulated in a comoving volume $V$, assumed to be sufficiently large and comprise many causally disconnected patches of the universe, is given by
\begin{equation}
\label{grl}
S =\frac{2 \xi V}{\tau} \; , 
\end{equation}
where $\tau$ is the conformal time and $\xi$ is the so called area (scaling) parameter. The universality is manifested in the area parameter taking value $\xi \approx 1.2$ for a variety of decelerating early universe cosmologies with the EoS parameter $w={\cal O} (1)$. This greatly extends the observation of Ref.~\cite{Dankovsky:2025pjg} arguing that the scaling value $\xi \approx 1.2$ obtained assuming RD universe holds for arbitrary (unbiased) initial conditions. The universality is approximate, and the departure from $\xi \approx 1.2$ becomes visible, as one approaches Minkowski space by considering very large values of $w$, i.e., $w \gtrsim 10$. Partially this tendency is attributed to the richer substructure of the DW network in the form of small closed walls for large $w$, as it is discussed in more details in sec.~\ref{sec:scaling}. Nevertheless, closed walls give an insignificant contribution to the net wall area, and the parameter $\xi$ remains confined to the narrow range $1.2 \leq \xi \leq 1.4$.

Together with universality of $\xi$, the law~\eqref{grl} implies that the particle horizon $l=a\tau$ sets the DW correlation length. 
Indeed, one can rewrite Eq.~\eqref{grl} in terms of physical area $S_{ph}=a^2S$ and volume $V_{ph}=a^3 V$ as $S_{ph}=2\xi V_{ph}/l$. In this form it is manifest that the particle horizon\footnote{Throughout the paper under {\it particle horizon} we mean the one calculated for the epoch with given $w=const$, excluding the previous history.} rather than the Hubble rate plays a major role in DW evolution. This fact remains 
hidden if one restricts to DW evolution at RD stage, because the particle horizon coincides with the inverse Hubble rate in this case. 

DWs are intrinsically inhomogeneous objects moving with relativistic velocities, and thus they may serve as powerful sources of GWs~\cite{Hiramatsu:2013qaa, Ferreira:2023jbu, Gruber:2024pqh, Dankovsky:2024zvs}.
The particle horizon associated with the correlation length of the DW network determines its main phenomenological properties including GW characteristics. We indeed observe in sec.~\ref{sec:gw} that the peak frequency of GWs is given by the inverse particle horizon with a high accuracy. This property holds for a large variety of universe EoS, and we demonstrate this by numerically evaluating GWs from simulated networks of {\it evolving} DWs. 

We also discuss a phenomenologically more interesting case of GWs left after DWs annihilate in the primordial plasma. It is well-known that stable (constant tension) DWs are inconsistent with CMB properties, unless they fulfill the bound $\sigma_{wall} \lesssim (1~\mbox{MeV})^3$, where $\sigma_{wall}$ is the DW tension~\cite{Zeldovich:1974uw, Lazanu:2015fua}. With such a stringent constraint, DWs would not serve as powerful sources of GWs or particles. The annihilation of DWs is typically triggered by introducing a potential bias  explicitly breaking a discrete symmetry and lifting degeneracy between potential minima. We demonstrate analytically and numerically that shapes of GW spectra from annihilating DWs are sensitive to the cosmological EoS. The dependence on the EoS parameter $w$ is prominent both in the infrared (IR) and ultraviolet (UV) parts of the spectra. 
 %Dependence of the IR slope on the EoS parameter derived in sec.~\ref{sec:gwth} coincides with that of Refs.~\cite{Cai:2019cdl, Hook:2020phx} in the case $w<1$, 
 %but differs in the case of kination, $w=1$, and stiffer EoS. Furthermore, we demonstrate numerically in sec.~\ref{sec:gw} presence of the plateau in the UV part of the spectrum previously identified in Ref.. for the case of RD universe. This feature of the spectrum is sometimes attributed to the numerical artifact. However, we argue for the physical origin of the plateau. We observe that the plateau grows wider and taller, as one increases $w$, and eventually affects the frequency range, which is arguably not affected by possibly lattice artifacts. Nevertheless, observation-wise, one  the near peak part of GW spectrum does not exhibit a strong dependence on $w$; in this regard,  Though 
However, the near peak part of the spectrum, which is perhaps most interesting observationally, rather weakly depends on $w$. We run high resolution numerical simulations to reconstruct GW spectra from annihilating DWs  
for a selection of primordial cosmologies, and perform an accurate fitting of GW spectra in the near peak region. In particular, we revisit GWs emitted by annihilating DWs during RD~\cite{Kitajima:2023cek, Cyr:2025nzf, Notari:2025kqq, Babichev:2025stm, Barbini:2026edx, Ferreira:2024eru}, for which purpose 
 we use the lattice with $4096^3$ grid points.

The paper is organized as follows. In sec.~\ref{sec:preliminaries} we briefly review basics of DW evolution and setup the scalar field system, where the DW network arises, for numerical simulations. In sec.~\ref{sec:scaling} we discuss results of simulations for DWs and demonstrate universality of the scaling property. We switch to the study of GWs produced by DWs in sec.~\ref{sec:gwth} and deduce relevant theoretical expressions assuming different cosmologies. In sec.~\ref{sec:gw} we show results of lattice simulations for GWs from stable and annihilating DWs. In the concluding section~\ref{sec:conclusions} we generalize the results regarding DW evolution to the case of an accelerated universe. There we also compare our findings with other works, where DWs in various cosmologies have been discussed.

%%%%%%%%%%%%%%%%%%%%%%%%%%%%%%%%%%%%%%%%%%%%%%%%%%%%%%%%%%%%%%%%%%%%%%%%%%
\section{Preliminaries}
\label{sec:preliminaries}

We consider a model of a real scalar $\chi$ evolving in a double well potential described by the Lagrangian:
\begin{equation}
\label{base}
{\cal L}=\frac{1}{2} \partial^{\mu} \chi\partial_\mu \chi -\frac{\lambda}{4} (\chi^2 -v^2)^2-\epsilon \chi^3 \; .
\end{equation}
 For the constant $\epsilon \ll \lambda v$, this model has an approximate $Z_2$-symmetry spontaneously broken by the expectation value $\langle \chi\rangle=\pm v$. In this setup, one naturally expects formation of long-lived DWs described by the tension 
\begin{equation}
\sigma_{wall}=\frac{2\sqrt{2 \lambda} v^3}{3} \; .
\end{equation}
Namely, the universe is getting dissected into regions, where the field $\chi$ is residing in one of two minima. These regions are separated by the walls with the width given by
\begin{equation}
\delta_{wall} = \frac{2}{m_{\chi}} = \frac{\sqrt{2}}{\sqrt{\lambda} v} \; ,
\end{equation}
where $m_{\chi}=\sqrt{2 \lambda} v$ is the mass of $\chi$-particles in the broken phase. In what follows, we neglect interactions of the DW field with other fields present in the primordial plasma. 
See, e.g., Refs.~\cite{Blasi:2023sej, Filippov:2025dpb} including such interactions. 
Consistently with this, we ignore possible thermal corrections to the Lagrangian~\eqref{base}. In this situation the DW network is formed when the field $\chi$, initially kept at zero by the Hubble friction, starts rolling to the potential minima as the Hubble rate drops below $m_{\chi}$. 

Soon after formation the network enters a scaling regime, which corresponds to the self-similar evolution of DWs described by the correlation length stretching proportionally to the cosmic time $t$. The network is mainly represented by one long wall extending throughout the horizon, while the total area comprised by small closed walls is negligible. See Ref.~\cite{Dankovsky:2024zvs} and sec.~\ref{sec:scaling}, where we provide a firm quantitative support for this picture.

In this work we study both cases $\epsilon=0$ and $\epsilon \neq 0$. One typically assumes $\epsilon \neq 0$ in order to induce the instability in the DW network. This breaks degeneracy between two vacua, with the energy density in the false vacuum being lifted relative to the true one by $V_{bias} \approx 2\epsilon v^3$. The system eventually ends up being in the true vacuum, and the DW network ceases to exist. This picture of DW annihilation is confirmed by various numerical simulations, see, e.g., recent Refs.~\cite{Babichev:2025stm, Ferreira:2024eru}. Nevertheless, some details of the process are still unclear, in particular estimation of the annihilation time of DWs. Some recent simulations~\cite{Babichev:2025stm} indicate a functional dependence $t_{ann} \propto 1/V^{2/3}_{bias}$, while Ref.~\cite{Barbini:2026edx} rather favors the ``traditional" estimate $t_{ann} \sim \sigma_{wall}/V_{bias}$. We do not contribute to this debate in the present work.

The case $\epsilon =0$, when exact $Z_2$-symmetry is recovered, corresponds to "eternal" DWs, which come to dominate the universe, unless their tension is chosen to be very low. That scenario is unacceptable. However, one can view $\epsilon=0$ as a viable approximation valid during a large portion of DW evolution. This is justified, since phenomenologically interesting values of $\epsilon$ are typically tiny, $\epsilon \ll \lambda v$. There remains a plausible option that the DW network dominated cosmic expansion for some period of primordial history. We leave this interesting scenario out of the scope of the present work and assume that DWs are strictly sub-dominant at all the times. See Refs.~\cite{Friedland:2002qs, Zeng:2026rxt} studying the DW dominated universe.

 In the universe governed by a fluid with the energy density evolving as in Eq.~\eqref{1st}, the scale factor grows with the conformal time $\tau$ as
\begin{equation}
a (\tau)=a_i \cdot \left(\frac{\tau}{\tau_i} \right)^{\frac{2}{1+3w}} \; .
\end{equation}
Note that for $w \gg 1$ the scale factor $a(\tau)$ is approximately constant, and we effectively recover Minkowski space.
The equation of motion following from the Lagrangian~\eqref{base} reads
\begin{equation}
\chi''+\frac{4}{(1+3w) \tau} \chi'-\partial^2_i \chi+\left(\frac{[1+3w] \cdot \tau }{2} \right)^{\frac{4}{1+3w}} \left[ \chi (\chi^2-1)+3\epsilon \chi^2 \right] =0 \; , 
\end{equation} 
where the primes stand for derivatives with respect to $\tau$. Hereafter we work with dimensionless variables and model constants:
\begin{equation}
x_i \rightarrow \sqrt{\lambda}v x_i\,, \qquad  k \rightarrow \frac{k}{\sqrt{\lambda}v}\,, \qquad \tau \rightarrow \sqrt{\lambda} v \tau\,, \qquad \chi \rightarrow \frac{\chi}{v}\,, \qquad \epsilon \rightarrow \frac{\epsilon}{\lambda v} \; ,
\end{equation}
where $k$ are wavenumbers of Fourier modes of the DW field $\chi$,
\begin{equation}
\chi ({\bf k}) =\int d{\bf x} e^{-i{\bf kx}} \chi ({\bf x}) \; .
\end{equation}
We set the initial scale factor $a_i=1$ and conformal time $\tau_i =\frac{2}{1+3w}$. This choice corresponds to the condition $H_i =\sqrt{\lambda} v$ being fulfilled at the onset of simulations. 

Let us define the optimal comoving box size $L$ for simulating the DW network on the lattice with grid number $N$. For this purpose, it is convenient to define the parameters
\begin{equation}
\label{kappas}
\kappa \equiv \frac{\delta_{wall} N}{ a(\tau_{max}) L}\,, \qquad \kappa' \equiv \frac{a(\tau_{max}) L}{2 H^{-1} (\tau_{max})} \; ,
\end{equation}
where $\tau_{max}$ is the maximally allowed conformal time of simulations of the DW network. Generically, this can be different from the actual final conformal time of simulations $\tau_f$. One imposes the conditions $\kappa \gtrsim 1$ and $\kappa' \gtrsim 1$, which ensure that the DW width always exceeds the lattice spacing, and that the DW network does not escape from the simulation box by the end of simulations. In terms of the lattice grid number $N$ and parameters $\kappa$ and $\kappa'$ the simulation time $\tau_{max}$ is expressed as 
\begin{equation}
\label{tauf}
\tau_{max} =\frac{2^{\frac{5+3w}{6(1+w)}}}{1+3w} \times \left(\frac{N}{\kappa \kappa'} \right)^{\frac{1+3w}{3(1+w)}} \; ,
\end{equation}
while the comoving box size $L$ is given by 
\begin{equation}
\label{L}
L= (2\kappa')^{\frac{2}{3(1+w)}} \times \left(\frac{\sqrt{2} N}{\kappa} \right)^{\frac{1+3w}{3(1+w)}}  \; .
\end{equation}
It is also instructive to write the ratio
\begin{equation}
\frac{\tau_{max}}{\tau_i}= \left(\frac{N}{\sqrt{2} \kappa \kappa'} \right)^{\frac{1+3w}{3(1+w)}} \; ,
\end{equation}
from which it is clear that the time span of simulations effectively reduces for smaller $w$, and increasing the grid number $N$ barely improves the situation. This is the reason, why we focus on the cases $w \geq 0$ in what follows. 

As in Ref.~\cite{Dankovsky:2025pjg}, we mostly set $\kappa=1$ and $\kappa'=\pi/2$, when studying evolution of DWs. However, we will assume other choices of $\kappa$ and $\kappa'$ depending on the problem at hand. For example, choosing different sets of $\kappa$ and $\kappa'$ has been proven essential for the detailed study of GW spectra, as it is discussed in Refs.~\cite{Dankovsky:2025pjg} and later in this paper. Second, the scale factor is growing very slowly in the case $w \gg 1$, and hence the condition $\delta_{wall} > a(\tau) L/N$ is only marginally fulfilled 
for the choice $\kappa=1$ according to Eq.~\eqref{kappas}. Therefore, we typically set $\kappa=5$ and $\kappa'=\pi/2$ when dealing with $w \gg 1$ .

Two important comments are in order here. First, $\tau_{max}$ is defined as the maximal time of simulations, while the DW network is present in the lattice box. If the DW network is annihilated in one or another way at some $\tau<\tau_{max}$, 
it becomes possible to extend the actual final conformal time of simulations $\tau_f$ beyond $\tau_{max}$, i.e., set $\tau_f \gg \tau_{max}$. Second, when defining the optimal lattice box parameters, we have been comparing the comoving box size with  $ (a H)^{-1}$. Nevertheless, it can be more appropriate to compare $L$  with the comoving particle horizon. This has a modest impact on the estimation of the optimal lattice box size, which differs from Eq.~\eqref{L} by the order unity factor $[2/(1+3w)]^{2/3(1+w)}$, and we continue working with Eq.~\eqref{L}. However, the allowed time of simulations is considerably altered:
\begin{equation}
\label{tauflonger}
\tau_{max} =\left(\frac{2}{1+3w} \right)^{\frac{2}{3(1+w)}}  \times \left(\frac{N}{\sqrt{2} \kappa \kappa'} \right)^{\frac{1+3w}{3(1+w)}} \; .
\end{equation}
It significantly exceeds Eq.~\eqref{tauf} for $w \gg 1$, and hence one can allow for a longer simulation run in this case.

Initial conditions for the scalar field $\chi$ are phrased via 2-point correlation functions of its Fourier mode $\chi ({\bf k})$ and the time derivative of the latter: 
\begin{equation}
\langle \chi ({\bf k}) \chi ({\bf q}) \rangle= (2\pi)^3 A(k) \delta ({\bf k}+{\bf q})\,, \qquad \langle \chi' ({\bf k}) \chi' ({\bf q}) \rangle= (2\pi)^3 B(k) \delta ({\bf k}+{\bf q}) \; .
\end{equation}
The scaling evolution of the DW network is expected to carry no memory of initial conditions, hence the functions $A(k)$ and $B(k)$ can be arbitrary. 
However, in realistic DW simulations on a lattice initial conditions matter~\cite{Dankovsky:2025pjg}. The likely reason is that initial scalar field configurations with enhanced infrared properties are more susceptible to non-physical effects related to a finite box size. A clear way to mitigate such effects is by choosing initial conditions with suppressed IR modes. This explains a somewhat peculiar choice of functions $A$ and $B$ adopted in our analysis~\cite{Dankovsky:2025pjg}:
\begin{equation}
\label{initial}
A(k)=\frac{1}{k} \cdot \left(e^\frac{k}{T}-1 \right)^3 \cdot \theta \left(1-k \right)\,, \qquad 
B(k)=k \cdot \left(e^\frac{k}{T}-1 \right)^3 \cdot \theta \left(1-k \right) \; ,
\end{equation}
where the constant $T$ is set at $T \approx 1.7$. Such initial conditions imply a sufficiently strong suppression in the IR. We have also imposed the UV cutoff with the theta-function: this is a typical prescription to avoid overly large scalar field inhomogeneities of the non-topological nature, which can be confused with DWs.

%%%%%%%%%%%%%%%%%%%%%%%%%%%%%%%%%%%%%%%%%%%%%%%
%%%%%%%%%%%%%%%%%%%%%%%%%%%%%%%%%%%%%%%%%%%%%%
%%%%%%%%%%%%%%%%%%%%%%%%%%%%%%%%%%%%%%%%%%%%%

\section{Universality of scaling} 
\label{sec:scaling}

\begin{figure}[!htb]
\begin{center}
 \includegraphics[width=0.99\textwidth]{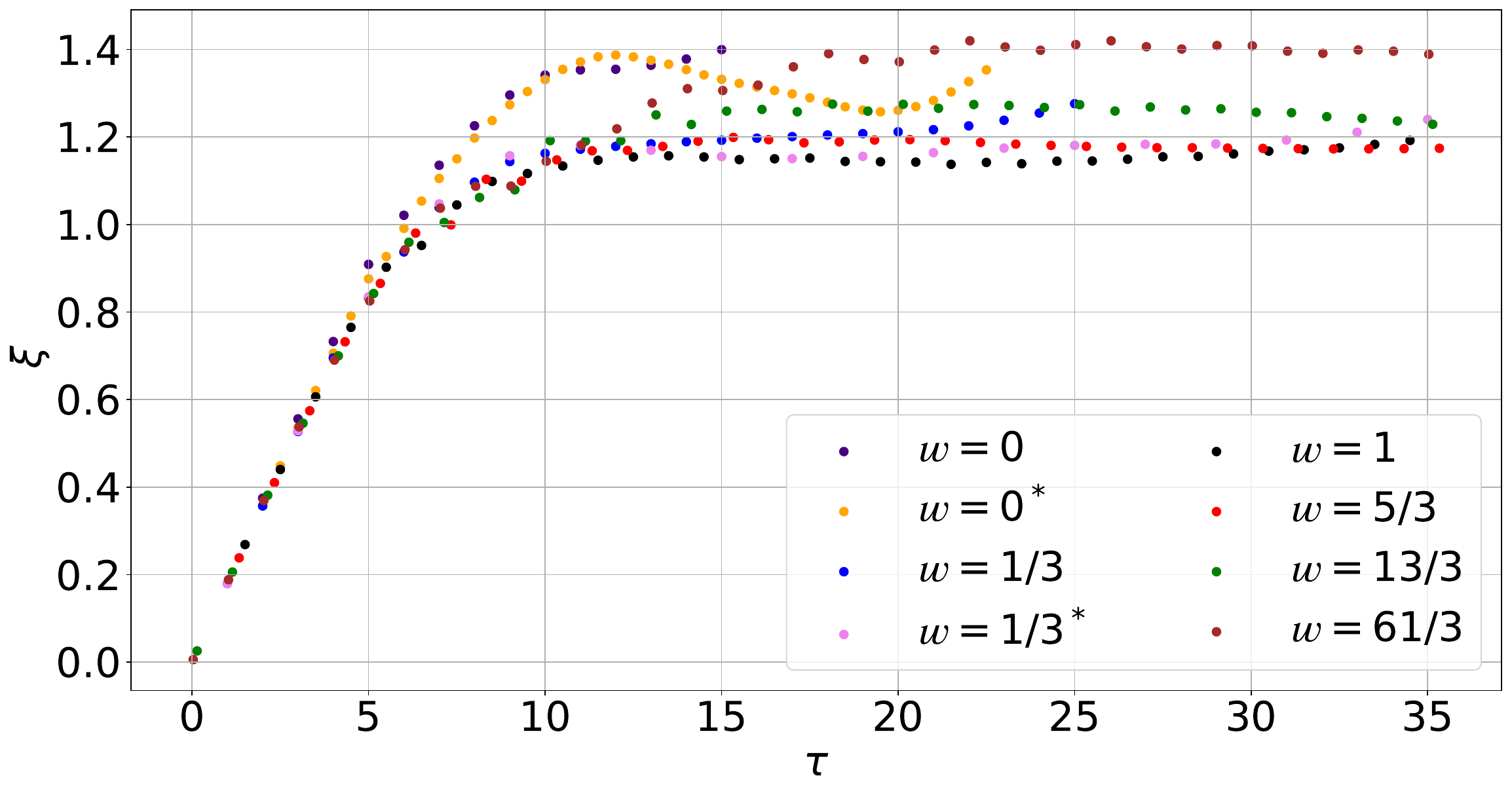} 
\end{center}
    \caption{The area parameter $\xi$ defined in Eq.~\eqref{universality} is demonstrated for different cosmologies characterized by the constant EoS parameter $w$. The simulations have been carried out with $1024^3$ lattice, except for the cases marked with an asterisk, where the $2048^3$ lattice has been used.} \label{scaling} 
\end{figure}

In this section we numerically evaluate DW area assuming various primordial cosmologies, i.e., various values of the EoS parameter $w$. For this purpose, we make use of the estimator developed in Ref.~\cite{Press:1989yh}: 
\begin{equation}
\label{estimator}
S=(\Delta x)^2 \sum \delta \frac{|\nabla \chi |}{|\partial_x \chi|+|\partial_y \chi|+|\partial_z \chi|} \; ,
\end{equation}
where the summation is over the pairs of lattice grid points; $\Delta x$ is the comoving distance between two neighboring lattice sites. One sets $\delta =1$, if the scalar $\chi$ changes its sign at the grid points of the pair, and otherwise $\delta=0$. As it is discussed in Refs.~\cite{Scherrer:1997sq, Li:2025gld}, 
applying the estimator~\eqref{estimator} one may overestimate the DW area by the factor $3/2$. 
As we are primarily interested in the time evolution of the DW area and how it changes depending on the background cosmology, this possible artificial amplification factor will not bother us in what follows. 

 We have performed the simulations using the $1024^3$ and $2048^3$ lattices with the comoving box size given by Eq.~\eqref{L}. We choose $\kappa=5$ for large $w=13/3$ and $w=61/3$; in all the other cases we choose $\kappa=1$. Another parameter is fixed to be $\kappa'=\pi/2$. See sec.~\ref{sec:preliminaries} for clarifications. 
It is convenient to present results of simulations in terms of the dimensionless area (scaling) parameter $\xi$:
\begin{equation}
\label{universality}
\xi \equiv \frac{S \tau}{2V} \; .
\end{equation}
 One can see from Fig.~\ref{scaling} that the area parameter reaches a constant value exactly meaning that DWs enter the scaling regime. The values taken by the parameter $\xi$ in the scaling regime are confined to a narrow range:
\begin{equation}
\label{approxi}
1.2 \leq \xi \leq 1.4 \; .
\end{equation}
For most cosmologies of interest, one can further narrow down this range, and the area parameter can be approximated as $\xi \approx 1.2$. We observe slightly larger $\xi$ (still in the range~\eqref{approxi}) in the case of matter dominated (MD) universe with $w =0$; recall, however, that low $w$ require significantly higher lattice resolution. For the analogous reason, we do not consider cosmologies with $w<0$. Away from MD universe, visible departures from $\xi \approx 1.2$ take place only for the EoS parameter $w =61/3$, i.e., as one approaches Minkowski space. We discuss a possible physical reason for such a tendency in the end of the section.

As it follows, the area parameter $\xi$ exhibits a rather high degree of universality through different cosmologies. This universality of $\xi$ suggests that the particle horizon,
\begin{equation}
\label{ph}
l=a \tau=a \int^t_0 \frac{dt'}{a(t')}\,, 
\end{equation}
rather than 
the inverse Hubble rate is the relevant parameter governing DW evolution in scaling. The fact that the cosmic expansion is not crucial in the evolution of $\xi$ could be envisaged from the fact that DWs enter the scaling regime also in Minkowski space~\cite{Garagounis:2002kt, Hindmarsh:1996xv, Dankovsky:2024ipq}. It is important that we have defined the area parameter $\xi$ as in Eq.~\eqref{universality}. In the case of RD evolution, this definition matches another often used one $\xi =St/(a(t) V)$. However, these two are not equivalent beyond the case of RD, and the scaling universality is manifest only with the definition~\eqref{universality}. 

The universal value $\xi \approx 1.2$ observed for $1/3 \lesssim w \lesssim 1$ differs slightly from the one obtained in the case of vacuum initial conditions, $\xi \approx 0.85$~\cite{Dankovsky:2024zvs}, see also Refs.~\cite{Hiramatsu:2013qaa} and~\cite{Cyr:2025nzf}. This decrease of $\xi$ has been addressed in Ref.~\cite{Dankovsky:2025pjg}; it is likely to have a non-physical origin, and it reflects the interplay between long wavelength modes describing initial conditions and a finite lattice box. See the discussion in the end of sec.~\ref{sec:preliminaries}. With the choice~\eqref{initial} this non-physical effect is suppressed meaning that $\xi \approx 1.2$ is a genuine value of the area parameter reached independently of initial conditions in a real universe. 
Our present study extends this observation made in Ref.~\cite{Dankovsky:2025pjg} in the case of RD universe to a wider range of cosmologies.

One important comment is in order here. The particle horizon is defined as the light travel distance from the beginning of the universe till a given time $t$. It is clear, however, that the time of DW network formation $t_i$ matters when evolution of DWs is concerned, while the times $t \ll t_i$ are irrelevant. Hence the lower bound in the integral in Eq.~\eqref{ph} should be set at $t_i$. This makes no difference in the decelerating universe, because the integral in Eq.~\eqref{ph} is saturated on the upper bound. However, this can be important when extrapolating results of our numerical simulations to the case of accelerating universe, see sec.~\ref{sec:conclusions}.

The energy density of DW network 
with the comoving area $S$ is approximately given by 
\begin{equation}
\rho_{wall} \approx \frac{\gamma \sigma_{wall} S}{V a} \; ,
\end{equation}
where $\gamma$ denotes the Lorentz boost factor $\gamma =1/\sqrt{1-v^2_{rms}}$, and $v_{rms}$ is the root mean square (RMS) wall velocity. Using Eq.~\eqref{universality} and $l=a \tau$, one obtains
\begin{equation}
\label{rhowalll}
\rho_{wall} \approx \frac{2\xi \gamma \sigma_{wall}}{l} \; .
\end{equation}
The analysis of $v_{rms}$ depending on the EoS parameter $w$ can be found in Ref.~\cite{Martins:2016ois}. According to the Table I there, $v_{rms}$ increases from $0.33$ to $0.46$, as one is varying $w$ from $w=0$ to $w=7/3$. Correspondingly, the Lorentz boost factor changes in the range $1.06 \lesssim \gamma \lesssim 1.12$. Hence, the universality of DW evolution also holds, when the scaling property is phrased in terms of DW energy density. Interestingly, the discussion in Ref.~\cite{Martins:2016ois} also points to a certain universality; there it is found that DW evolution through different cosmologies can be described by the 6-parameter extension of the VOS model.

\begin{figure}[!htb]
\begin{center}
  \includegraphics[width=0.995\textwidth]{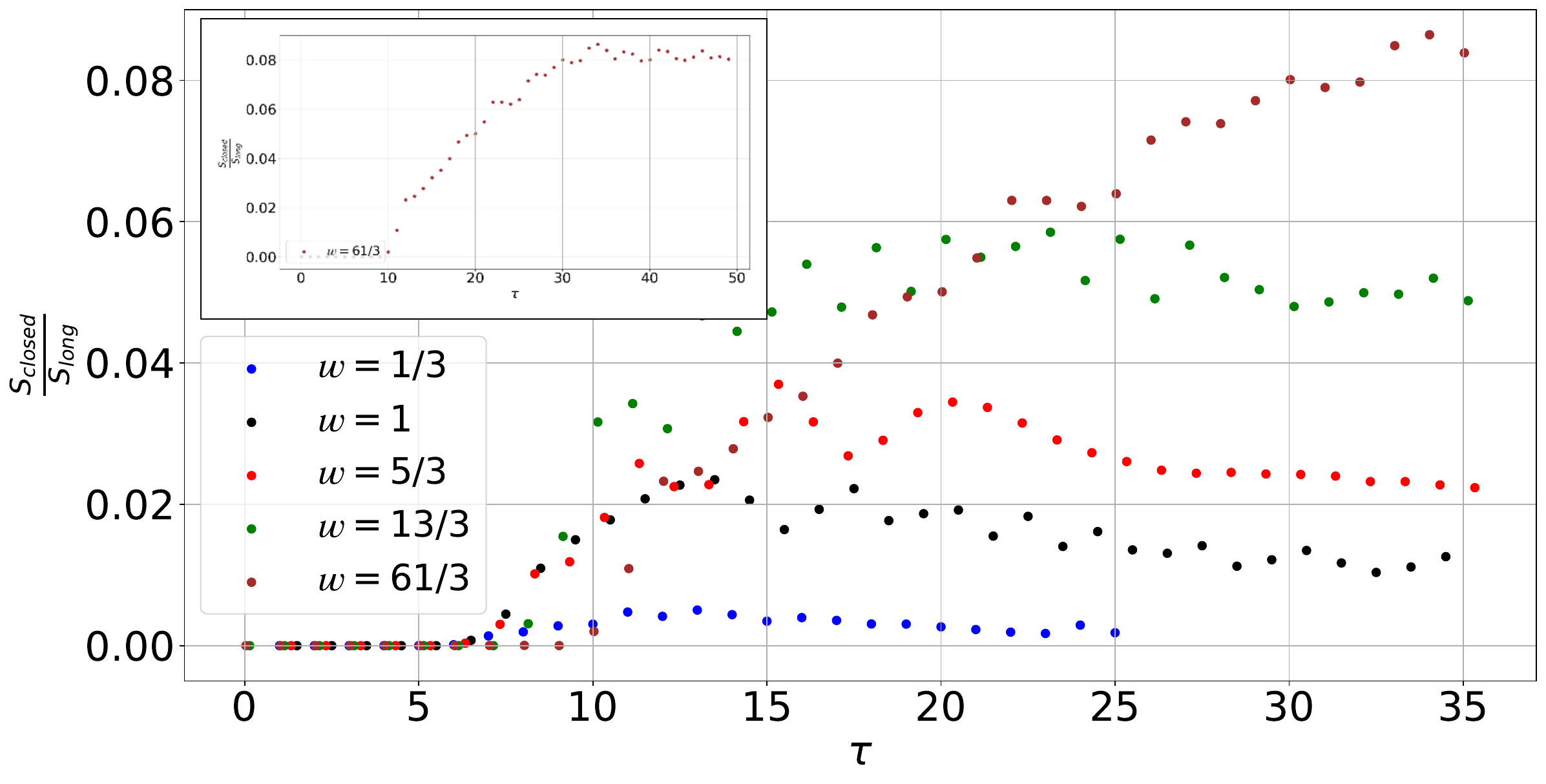} 
\end{center}
    \caption{The ratio of closed walls to the long wall surface areas is demonstrated assuming different cosmologies characterized by a constant EoS parameter $w$.} \label{closed}
\end{figure}

Related to this, let us note that the DW correlation length is often directly defined via the DW energy density~\cite{Martins:2016ois}:
\begin{equation}
\label{lcorr}
l_{e} \equiv \frac{\sigma_{wall}}{\rho_{wall}} \; .
\end{equation} 
(Here the subscript $e$ stresses the relation of so defined correlation length $l_e$ to DW energy density.) 
The ratio $l_{e}/l$ (which corresponds to $\xi_c/\tau$ in notations of Ref.~\cite{Martins:2016ois}) has been studied in Ref.~\cite{Martins:2016ois} for different values of $w$. Results demonstrated in Table I there exhibit a remarkable regularity, i.e., the ratio $l_e/l$ is quite well centered around $l_{e}/l \approx 0.5$ in a rather wide range of values of $w$. This agrees well with our results, because the ratio $l_{e}/l$ is related to the area parameter $\xi$ as 
\begin{equation}
\frac{l_{e}}{l}=\frac{1}{2\xi \gamma} \; .
\end{equation}
Hence, universality of $\xi$ should be reflected in the ratio $l_{e}/l$. Note, however, that the values $\xi \approx 1.2$ inferred from Fig.~\ref{scaling} yield slightly smaller $l_{e}/l$ compared to Ref.~\cite{Martins:2016ois}. That discrepancy may be related to the aforementioned choice of initial conditions and/or to the PRS prescription used in Ref.~\cite{Martins:2016ois}, and it deserves a more thorough investigation in the future.

Now we address a slight violation of universality observed in Fig.~\ref{scaling} for large $w$. This violation is likely to be attributed to the larger amplitudes of waves propagating on a long wall and the related enhancement of closed wall formation in the case $w \gg 1$. Let us explain this point. Waves moving on a wall may end up with a shock formation potentially leading to the detachment of small closed walls~\cite{Babichev:2026zca, Blanco-Pillado:2025gzs}. The Hubble friction damps the wave amplitude counteracting shock formation. By increasing $w$ one softens effects of cosmic expansion allowing for a larger amplitude of waves propagating on the wall thus making the wall more curved. This also triggers a more abundant formation of closed walls through shocks. Consequently, we get a larger area parameter $\xi$ upon increasing $w$ in accordance with Fig.~\ref{scaling}.

We support this heuristic picture with Fig.~\ref{closed}, where it is demonstrated that the total area of closed walls indeed grows with the increase of $w$. We see in particular that $\sim 10\%$ increase in $\xi$ observed in Fig.~\ref{scaling} for $w=61/3$ (compared to the cases with $w={\cal O} (1)$) matches $\sim 10\%$ fraction of closed walls seen for the same $w$, as it follows from Fig.~\ref{closed}. The algorithm used to evaluate the number and areas of closed DWs is rather straightforward. First we establish all the points corresponding to DWs. Then, we iterate once again over the lattice, find the first point where the DW network exists, and assign number
``1'' to the corresponding DW. Since DWs are continuous, we can establish all the points on the lattice constituting this DW. After this we continue iterating over the lattice until we find a point which belongs to the DW network, but not the 1st wall established in the previous iteration. We assign number ``2'' to this DW, and then repeat the procedure again, until the whole lattice is covered. At the end of the procedure we get the exact number of DWs, and we can calculate the area of each DW with the use of the estimator~\eqref{estimator}. 

While the fraction of closed walls grows with the increase of $w$ according to Fig.~\ref{closed}, they still give a small and often negligible (for $1/3 \lesssim w \lesssim 1$) contribution 
to the total area of the network, which is largely dominated by the long wall~\cite{Dankovsky:2024zvs}. In agreement with this, the best fit extended VOS model proposed in Ref.~\cite{Martins:2016ois} favors the near zero value of the so called chopping parameter $c_w$ responsible for production of closed walls. This explains why the universality of scaling is only slightly broken. Finally, it is worth mentioning that the value $\xi \approx 1.4$ observed in Fig.~\ref{scaling} for large $w =61/3$, when Minkowski space is effectively recovered, is also observed in the case of so called melting DWs~\cite{Dankovsky:2024ipq, Vilenkin:1981zs, Ramazanov:2021eya, Babichev:2021uvl, Babichev:2023pbf} characterized by a temperature dependent tension decreasing with cosmic expansion. Evolution of such melting walls in the RD universe is equivalent to evolution of conventional walls in the Minkowski space~\cite{Dankovsky:2024ipq, Babichev:2021uvl}. This consistency with the case of melting DWs serves as an additional cross-check of our simulations.

%%%%%%%%%%%%%%%%%%%%%%%%%%%%%%%%%%%%%%%%%%%%%%%%%%%%%%%%%%%%%%
\section{Gravitational waves: theoretical expectations}
\label{sec:gwth}

Before delving into details of numerical simulations of GWs from DWs, it is instructive to discuss theoretical expectations on this subject. The expressions presented below are valid for generic primordial sources, unless the case of DWs is specified. 
Assuming spatially flat FLRW Universe, we write the spacetime interval as follows: 
\begin{equation}
ds^2=a^2 (\tau) \left[d\tau^2-(\delta_{ij}+h_{ij}(\tau,x)) dx^i dx^j \right] \; .
\end{equation}
We neglect scalar metric perturbations hereafter and focus on the transverse traceless (TT) tensor perturbations $h_{ij}$ constituting GWs. The latter are sourced by the TT part of the source stress-energy tensor $T^{TT}_{ij} \equiv a^2 \Pi_{ij}$. The energy density of GWs produced by DWs is obtained upon averaging over many GW wavelengths and periods: 
\begin{equation}
\label{GW-def}
\rho_{gw}=\frac{1}{32\pi G_N a^2 (\tau)} \left \langle \frac{\partial h_{ij}}{\partial \tau} \frac{\partial h_{ij}}{\partial \tau} \right \rangle \; ,
\end{equation}
where $G_N$ is the Newton's constant. It is convenient to switch to Fourier modes of the GW field $h_{ij}$ and the source $\Pi_{ij}$: 
\begin{equation}
h_{ij} ({\bf k}, \tau)= \int d {\bf x} e^{-i{\bf kx}}   h_{ij} ({\bf x}, \tau)\,,  \qquad \Pi_{ij} ({\bf k}, \tau)= \int d {\bf x} e^{-i{\bf kx}}   \Pi_{ij} ({\bf x}, \tau)  \; .
\end{equation}
One also performs splitting of the GW field and the source into polarizations: 
\begin{equation}
h_{ij} ({\bf k}, \tau) =\sum_A h_A ({\bf k}, \tau) e^{A}_{ij} ({\bf k})\,, \qquad \Pi_{ij} ({\bf k}, \tau) =\sum_A \Pi_A ({\bf k}, \tau) e^{A}_{ij} ({\bf k}) \; ,
\end{equation}
where $e^A_{ij}$ denotes a pair of polarization tensors, $A=1,2$, satisfying the normalization condition $\sum_{i,j} e^{A}_{ij} e^{A'}_{ij}=2\delta_{AA'}$.

The equation of motion for the GW amplitude takes a convenient form in terms of the rescaled variable $\tilde{h}_A=a h_A$~\cite{Maggiore}:
\begin{equation}
\tilde{h}''_A ({\bf k}, \tau)+\left(k^2-\frac{a''}{a} \right) \tilde{h}_A ({\bf k}, \tau) =16\pi G_N a^3 (\tau) \Pi_{A} ({\bf k}, \tau) \; .
\end{equation}
Its solution satisfying initial conditions $h ({\bf k}, \tau)|_{\tau=\tau_i}=0$ and $h' ({\bf k}, \tau) |_{\tau=\tau_i}=0$ reads
\begin{equation}
\label{solu}
h_A ({\bf k}, \tau) =\frac{16 \pi  G_N}{k a(\tau)} \int^{\tau}_{\tau_i} d\tilde{\tau} \left(\frac{\tau}{\tilde{\tau}} \right)^{1/2} a^3 (\tilde{\tau}) \frac{J_{\nu} (k\tilde{\tau}) Y_{\nu} (k\tau) -J_{\nu} (k\tau) Y_{\nu} (k\tilde{\tau})}{W(k \tilde{\tau})} \cdot \Pi_A ({\bf k}, \tilde{\tau}) \; .
\end{equation}
where $J_{\nu} (k\tau)$ and $Y_{\nu} (k\tau)$ are the Bessel functions of the first and second kind, respectively, of the order
\begin{equation}
\label{nu}
\nu =\frac{|3-3w|}{2|1+3w|} \; ,
\end{equation}
and $W (k \tau)$ is the Wronskian given by  
\begin{equation}
W (k\tilde{\tau})=J_{\nu} (k\tilde{\tau})  \frac{\partial Y_{\nu} (k\tilde{\tau})}{\partial (k\tilde{\tau})}-Y_{\nu} (k\tilde{\tau}) \frac{\partial J_{\nu} (k\tilde{\tau})}{\partial (k\tilde{\tau})} \; .
\end{equation}
We assume that all the relevant stages of evolution of interesting GW modes, from their production till horizon entrance (if superhorizon at production), take place during one and the same epoch described by the constant EoS parameter $w$.
When $w=1/3$, one has $\nu=1/2$, in which case $J_{\nu} (k\tau) =\sqrt{\frac{2}{\pi k\tau}} \sin k\tau$ and $Y_{\nu} (k\tau)=-\sqrt{\frac{2}{\pi k\tau}} \cos k\tau$, and we recover from Eq.~\eqref{solu} the correct solution in the RD universe, see, e.g., Eq. (19) of Ref.~\cite{Dankovsky:2024zvs}:
\begin{equation}
h_A ({\bf k}, \tau) =\frac{16 \pi  G_N}{k a(\tau)} \int^{\tau}_{\tau_i} d\tilde{\tau} a^3 (\tilde{\tau}) \sin k[\tau-\tilde{\tau}] \cdot \Pi_A ({\bf k}, \tilde{\tau}) \; .
\end{equation}
 Next we use isotropy and homogeneity of the spacetime as well as no preference for a particular GW polarization and write down
\begin{equation}
\label{correlator}
\langle \Pi_A ({\bf k}, \tau_1) \Pi_B ({\bf q}, \tau_2) \rangle=(2\pi)^3 \delta ({\bf k}+{\bf q}) \delta_{AB} \rho_{s} (\tau_1) \rho_{s} (\tau_2) P(k, \tau_1, \tau_2) \; ,
\end{equation}
where $P(k, \tau_1, \tau_2)$ is the source power spectrum, and we have factored out the source energy density $\rho_s$ averaged over the space.

From Eq.~\eqref{GW-def} we obtain for the spectral energy density per logarithmic momentum: 
\begin{equation}
\label{spectrumth}
\begin{split}
&\frac{d\rho_{gw}}{d\ln k} =\frac{16 G_N k}{\pi a^4 (\tau)} \cdot   \int^{\tau}_{\tau_i} \int^{\tau}_{\tau_i} d\tau_1 d\tau_2 a^3 (\tau_1) a^3(\tau_2) \cdot \left(\frac{\tau^2}{\tau_1 \tau_2} \right)^{1/2} \cdot \frac{ \rho_{s} (\tau_1)  \rho_{s} (\tau_2) P(k, \tau_1, \tau_2)}{W(k\tau_1) W(k\tau_2)}  \times \\ 
& \Bigl[\Bigl(J_{\nu} (k\tau_1) Y'_{\nu} (k\tau) -J'_{\nu} (k\tau) Y_{\nu} (k\tau_1) \Bigr) \cdot \Bigl(J_{\nu} (k\tau_2) Y'_{\nu} (k\tau) -J'_{\nu} (k\tau) Y_{\nu} (k\tau_2) \Bigr)+ \\ 
&+\Bigl(\frac{1}{\tau}-2{\cal H} \Bigr) \cdot \Bigl(J_{\nu} (k\tau_1) Y'_{\nu} (k\tau) -J'_{\nu} (k\tau) Y_{\nu} (k\tau_1) \Bigr) \cdot \Bigl(J_{\nu} (k\tau_2) Y_{\nu} (k\tau) -J_{\nu} (k\tau) Y_{\nu} (k\tau_2) \Bigr)+\\
&+\Bigl(\frac{1}{2\tau}-{\cal H} \Bigr)^2 \cdot \Bigl(J_{\nu} (k\tau_1) Y_{\nu} (k\tau) -J_{\nu} (k\tau) Y_{\nu} (k\tau_1) \Bigr) \cdot \Bigl(J_{\nu} (k\tau_2) Y_{\nu} (k\tau) -J_{\nu} (k\tau) Y_{\nu} (k\tau_2) \Bigr)\Bigr]\; .
\end{split}
\end{equation}
This expression exhibits an explicit dependence on $\nu$ and hence on the universe EoS through the Bessel functions $J_{\nu}$ and $Y_{\nu}$. We can considerably simplify the expression for the GW spectrum $d\rho_{gw}/d\ln k$ by focusing on the modes obeying $k\tau \gg 1$. This is definitely a very good approximation from the viewpoint of current searches for the GW stochastic background. In this case, we can safely drop the last two terms in the square brackets in Eq.~\eqref{spectrumth} and obtain
\begin{equation}
\label{shortened}
\begin{split}
&\frac{d\rho_{gw}}{d\ln k} \approx \frac{16 G_N k}{\pi a^4 (\tau)} \cdot   \int^{\tau_e}_{\tau_i} \int^{\tau_e}_{\tau_i} d\tau_1 d\tau_2 a^3 (\tau_1) a^3(\tau_2) \cdot \left(\frac{\tau^2}{\tau_1 \tau_2} \right)^{1/2} \cdot \frac{\rho_{s} (\tau_1) \rho_{s} (\tau_2)P(k, \tau_1, \tau_2)}{W(k\tau_1) W(k\tau_2)}  \times \\ 
& \times   \Bigl(J_{\nu} (k\tau_1) Y'_{\nu} (k\tau) -J'_{\nu} (k\tau) Y_{\nu} (k\tau_1) \Bigr) \cdot \Bigl(J_{\nu} (k\tau_2) Y'_{\nu} (k\tau) -J'_{\nu} (k\tau) Y_{\nu} (k\tau_2) \Bigr) \; ,
\end{split}
\end{equation}
where $\tau_e$ stands for the moment when the source terminates. Here we explicitly assume that the source of GWs has a finite duration in time, which is the case of realistic annihilating DWs. Note that $\tau_e$ is not to be confused with the final time of simulations $\tau_f$ and the DW annihilation time $\tau_{ann}$. Namely, production of GWs continues some time after DWs are being destroyed~\cite{Kitajima:2023cek, Cyr:2025nzf, Notari:2025kqq, Babichev:2025stm, Barbini:2026edx}, likely due to inhomogeneities of scalar radiation left upon their collapse. However, it is often plausible to ignore the difference between $\tau_e$ and $\tau_{ann}$, since these are of the same order of magnitude, i.e., $\tau_e \approx 1.8 \tau_{ann}$~\cite{Dankovsky:2025pjg}.

The power spectrum $P(k, \tau_1, \tau_2)$ is independent of the momentum $k$ for superhorizon modes, $k \tau_n \ll 1$, where $n=1,2$. This universal property following from causality considerations~\cite{Caprini:2009fx, Cai:2019cdl,Hook:2020phx} can be used to infer some generic properties of GW sources. 
For this purpose we focus on the modes being superhorizon by the time when the source terminates, i.e., $k \tau_{ann} \ll 1$ and hence $k \tau_n \ll 1$. In this limit we can use the well-known small-argument approximations for the Bessel functions:
\begin{equation}
\label{small}
J_{\nu} (k\tau_n) \approx \frac{1}{\Gamma (\nu+1)} \left(\frac{k\tau_n}{2} \right)^{\nu}\,,  \qquad  Y_{\nu} (k\tau_n) \approx -\frac{\Gamma (\nu)}{\pi} \left(\frac{2}{k\tau_n} \right)^{\nu} \; .
\end{equation}
At the same time, since we assume $k \tau \gg 1$, one exploits the following large-argument approximations:
\begin{equation}
\label{large}
J_{\nu} (k\tau) \approx \sqrt{\frac{2}{\pi k\tau}} \cos \left(k\tau-\frac{\pi \nu}{2}-\frac{\pi}{4} \right)\,, \qquad Y_{\nu} (k\tau) \approx \sqrt{\frac{2}{\pi k\tau}} \sin \left(k\tau-\frac{\pi \nu}{2}-\frac{\pi}{4} \right) \; .
\end{equation}
Substituting Eqs.~\eqref{small} and~\eqref{large} into Eq.~\eqref{shortened} and recalling that $P(k, \tau_1, \tau_2)$ is independent of $k$, we infer the spectral shape of GWs in the IR limit:
\begin{equation}
\label{spectrumir}
\frac{d\rho_{gw}}{d\ln k} \propto k^{4-2\nu}\,,  \qquad k\tau \gg 1\,, \quad k\tau_{ann} \ll 1\,, \quad w \neq 1 \; .
\end{equation}
 This result is important, because it demonstrates that universality with respect to different cosmologies is broken at the level of GW spectra. In particular, the IR spectral index $n_{IR}=4-2\nu$ takes the value $n_{IR}=3$ in the RD universe~\cite{Caprini:2009fx}. The case of kination, $w=1$, is special, since it corresponds to $\nu=0$, and the Bessel functions are approximated as $J_{0} (k\tau_n) \approx 1$ and $Y_{0} (k\tau_n) \approx (2/\pi) \ln (k\tau_n/2)$. Substituting these into Eq.~\eqref{shortened} instead of Eq.~\eqref{small}, we get
\begin{equation}
\label{kinspectrum}
\frac{d\rho_{gw}}{d\ln k} \propto k^4 \ln^2 (k\tau_{ann}/2)\,, \qquad k\tau \gg 1\,, \quad k\tau_{ann} \ll 1\,, \quad w = 1 \; .
\end{equation}
(Recall that $\tau_{e} \sim \tau_{ann}$). We observe the double-logarithm amplification compared to what could be naively expected from Eq.~\eqref{spectrumir}.

Production of GWs in cosmologies different from RD has been previously studied in the literature. The result obtained in Refs.~\cite{Cai:2019cdl,Hook:2020phx} (see also Ref.~\cite{Blasi:2025tmn}) for the IR spectral index reads $n_{IR}=(1+15w)/(1+3w)$. This agrees with the spectral index in Eq.~\eqref{spectrumir} for $w<1$, but there is a growing discrepancy for $w \geq 1$, which starts with the case of kination, where we have identified the double-logarithmic factor missed previously. Furthermore, in the limit $w \rightarrow \infty$ we end up with $d\rho_{gw}/d\ln k \propto k^3$, as it should be in the case of Minkowski space, while the behavior $d\rho_{gw}/d\ln k \propto k^5$ follows from Refs.~\cite{Cai:2019cdl, Hook:2020phx}.

The limit $k \tau \ll 1$ is less interesting from the viewpoint of realistic GW observations, but it is relevant for lattice simulations of stable unbiased DWs. In that case one should use Eq.~\eqref{spectrumth} (where still $\tau\gg \tau_i$) and insert the small-argument expansion~\eqref{small} for 
all the Bessel functions. There we also substitute $\rho_{s} =\rho_{wall}$ and use Eq.~\eqref{rhowalll}. We assume the power spectrum $P(k, \tau_1, \tau_2)$ to be of the form~\cite{Blasi:2025tmn, Ramazanov:2023eau}:
\begin{equation}
\label{approx}
P(k, \tau_1, \tau_2) \approx C (\tau_1 \tau_2)^{3/2} \; ,
\end{equation}
where $C$ is a constant. This is the simplest possible expression, which fulfills the property of scaling, dimensionality of $P(k, \tau_1, \tau_2)$, and causality enforcing $P(k, \tau_1, \tau_2)$ to be independent of $k$ in the limit of interest $k \tau_n \ll 1$. Results of our previous sections suggest that the constant $C$ has little to no sensitivity to background cosmology. Performing integrations in Eq.~\eqref{spectrumth}, we obtain
\begin{equation}
\label{theor}
\frac{d\rho_{gw}}{d\ln k} =\frac{256 CG_N \xi^2 \gamma^2 \sigma^2_{wall}  (1+3w)^2 (k\tau)^3}{(15+ 9w)^2\pi} \; .
\end{equation}
One learns in particular that the IR spectral slope is described by 
\begin{equation}
\label{spectrumtr}
\frac{d\rho_{gw}}{d\ln k} \propto k^3
\end{equation}
independently of background cosmology. From the comparison with Eqs.~\eqref{spectrumir} and~\eqref{kinspectrum} we learn that extrapolating results obtained for stable DWs to the case of annihilating DWs is generally impossible. Only in the cases of RD universe and Minkowski space Eqs.~\eqref{spectrumir} and~\eqref{spectrumtr} give the same. 
We also learn from Eq.~\eqref{theor} that for the fixed $k$ and $\tau$, such that $k\tau \ll 1$, the spectrum grows with $w$ for $w>-1/3$ and eventually reaches saturation in the limit of large $w$. 
Since the IR slope is universal in the regime under consideration, it is tempting to extrapolate Eq.~\eqref{theor} to the peak associated with the comoving horizon size and compare with results of numerical simulations discussed in the next section.

%%%%%%%%%%%%%%%%%%%%%%%%%%%%%%%%%%%%%%%%%%%%%%%%%%%%%%%%%%
\section{Gravitational wave spectra: numerical analysis}

\label{sec:gw}

We switch to the numerical study of GWs from DWs in various early universe cosmologies. This is of considerable interest in view of ongoing and future searches for the stochastic GW background by PTA observatories~\cite{NANOGrav:2023gor, NANOGrav:2023hvm, EPTA:2023fyk, EPTA:2023xxk, Xu:2023wog, Reardon:2023gzh}, LISA~\cite{LISA:2017pwj}, Einstein Telescope (ET)~\cite{ET:2025xjr}, Cosmic Explorer~\cite{Reitze:2019iox}, and TianQin~\cite{Luo:2025ewp}. It is worth mentioning that recent PTA measurements point to the existence of stochastic GW background~\cite{NANOGrav:2023gor, EPTA:2023fyk, Xu:2023wog, Reardon:2023gzh}, and DWs can be the origin of this signal~\cite{NANOGrav:2023hvm}. As it is common, we describe GWs using the spectral fractional energy density per logarithmic frequency, or spectral energy density for short: 
\begin{equation}
\Omega_{gw} (k) \equiv \frac{1}{\rho_{U}} \cdot \frac{d \rho_{gw}}{d\ln k} \; .
\end{equation}
Recall that we assume the universe being filled with a barotropic fluid described by a constant EoS parameter,  
$w \equiv {\cal P}_U/\rho_U=\mbox{const}$.

\subsection{Near peak gravitational wave spectrum from stable domain walls}

\begin{figure}[!htb]
\begin{center}
    \includegraphics[width=\textwidth]{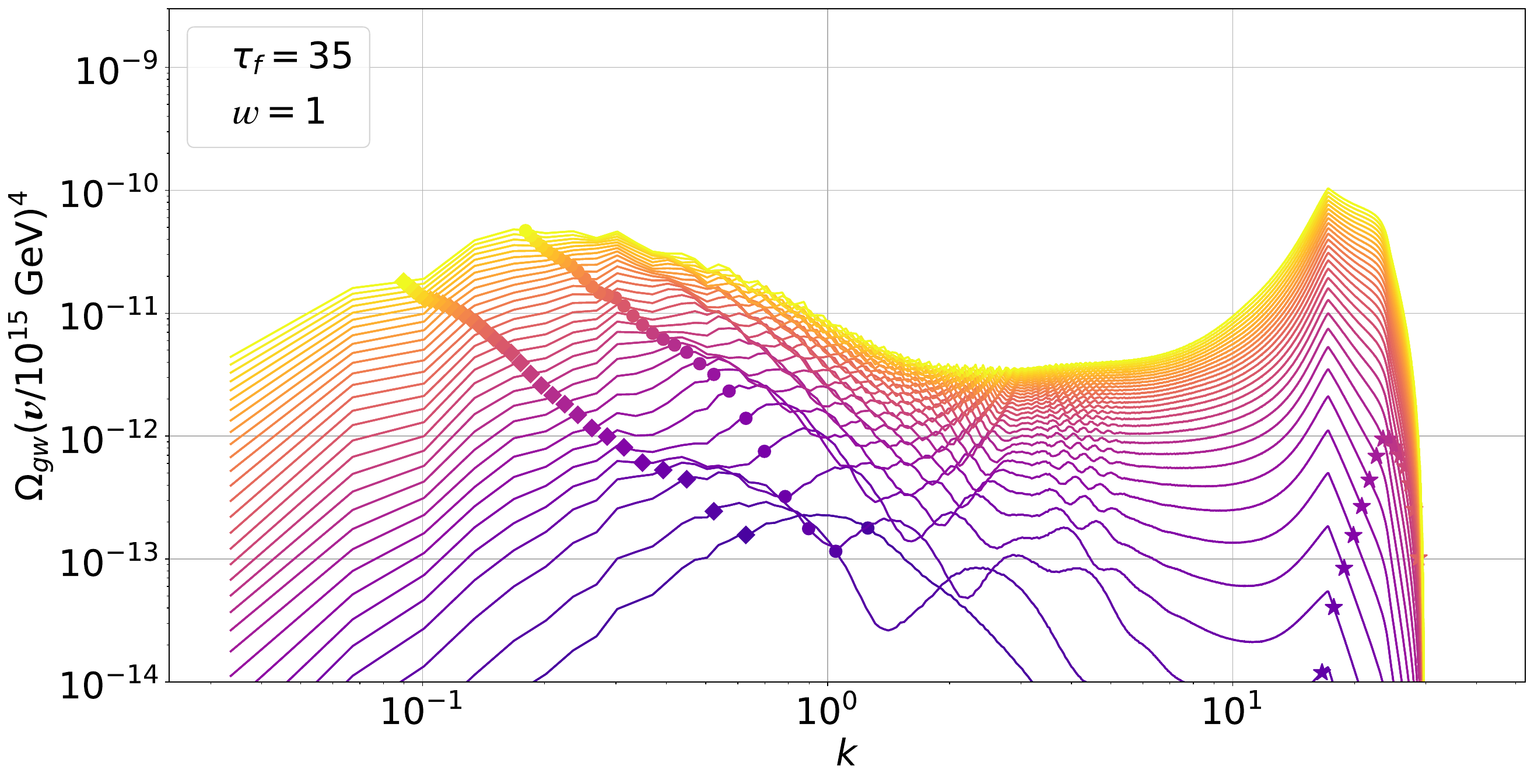} 
       \includegraphics[width=\textwidth]{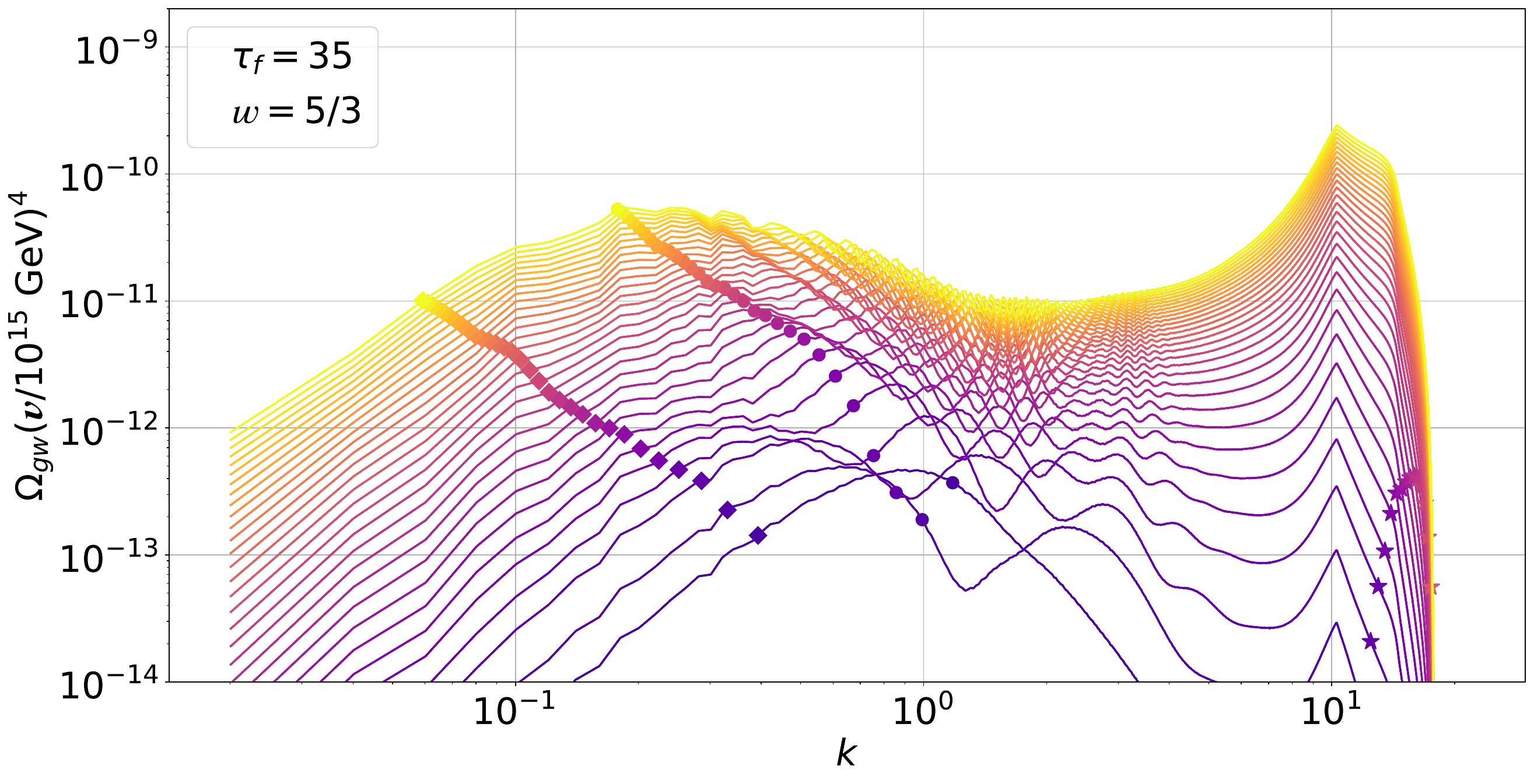} 
\end{center}
    \caption{GW spectra from stable DWs produced with $1024^3$ lattice assuming the universe EoS parameter $w=1$ (top panel) and $w=5/3$ (bottom panel). The squares, circles, and stars mark positions corresponding to the comoving Hubble rate, i.e., $k=2\pi a H$, to the particle horizon $k=2\pi/\tau$, and to the inverse DW width, i.e., $k=2\pi a/\delta_{wall}$, respectively.}\label{peaksVShubble} 
\end{figure}

\begin{figure}[!htb]
\begin{center}                \includegraphics[width=0.49\textwidth]{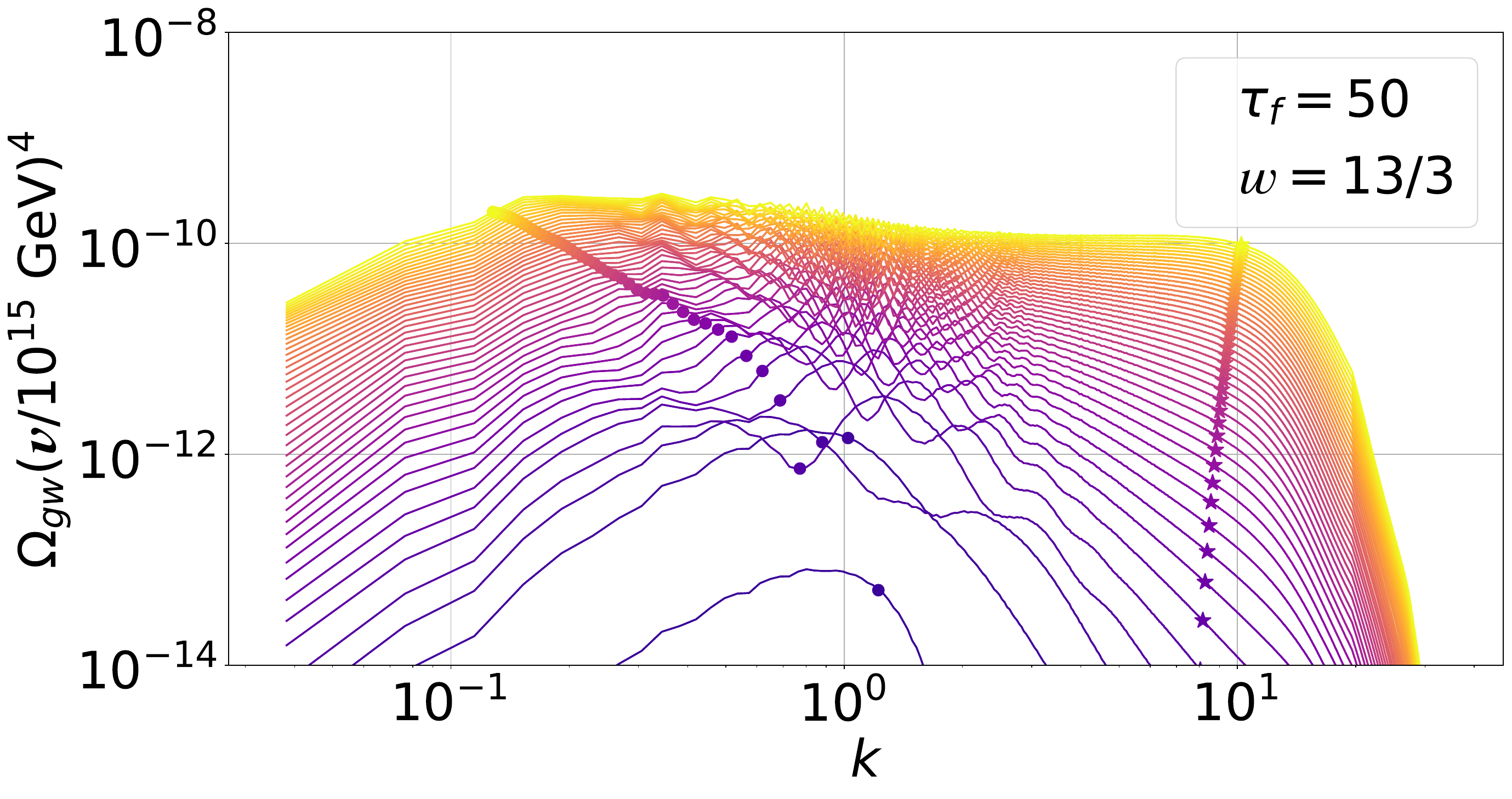} 
                           \includegraphics[width=0.49\textwidth]{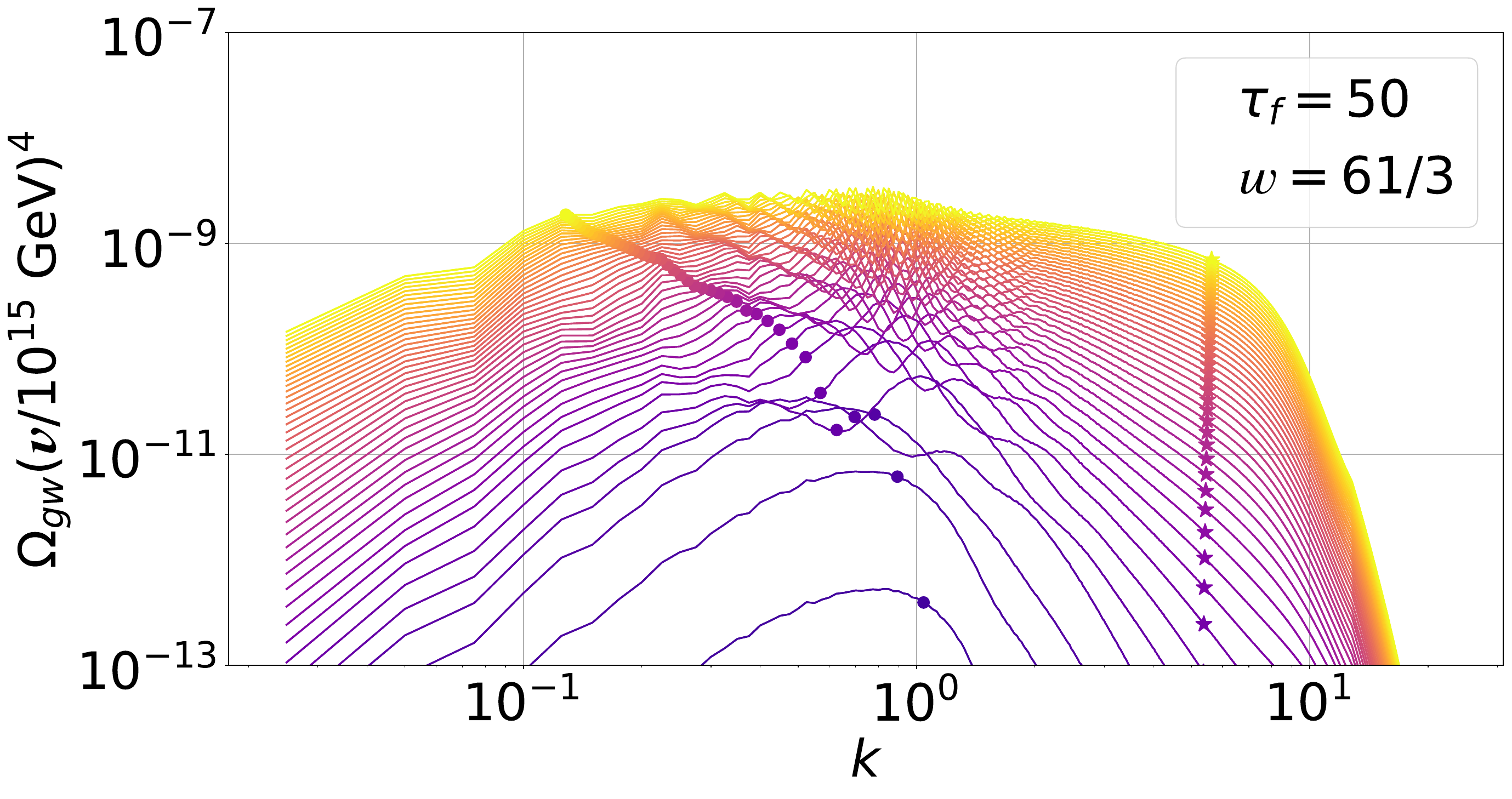} 
\end{center}
    \caption{GW spectra from stable DWs produced with $1024^3$ lattice assuming cosmologies with a very stiff EoS: $w=13/3$ (left panel), $w=61/3$ (right panel). The circles and stars mark positions corresponding to the particle horizon $k=2\pi/\tau$ and to the inverse DW width, i.e., $k=2\pi a/\delta_{wall}$, respectively.}\label{plateau} 
\end{figure}

We start with the analysis of GWs generated by evolving DWs. Namely, we assume that the observer living in the primordial universe described by the EoS parameter $w$ detects GWs from existing DWs. Albeit this case effectively corresponding to $\epsilon=0$ is less interesting 
phenomenologically, it is perfectly suitable for the demonstration of DW universality discussed in sec.~\ref{sec:scaling}. In this case we primarily focus on the vicinity of the peak, and therefore it is sufficient to use the lattice with $1024^3$ grid points. The results 
are shown in Figs.~\ref{peaksVShubble} and~\ref{plateau} for a selection of EoS parameters $w=(1, 5/3, 13/3, 61/3)$. In the cases $w=1$ and $w=5/3$, one observes a common twin peaks structure of GW spectra from DWs. The case of RD stage, $w=1/3$,  
is fairly similar and can be found in Fig.~8 of Ref.~\cite{Dankovsky:2025pjg}. In fact, only the left peak is physical, 
as advocated in Ref.~\cite{Dankovsky:2024zvs} and discussed in sec.~\ref{ann}. From Fig.~\ref{peaksVShubble} we infer that the physical GW peak is located at
\begin{equation}
\label{peakhorizon}
k_{peak} \approx \frac{2\pi}{\tau} \; ,
\end{equation}
at least for not very large $w$. This result is to be contrasted to the commonly assumed estimate for the peak $k_{peak} \approx 2\pi H a$. While these two expressions coincide  in the RD universe, there is a mismatch for $w \neq 1/3$, and Eq.~\eqref{peakhorizon} is clearly more accurate. We also find a very good agreement between Eq.~\eqref{peakhorizon} and the data obtained in\footnote{See Table 5 there, where one identifies $x_p \equiv k_{peak}/(2\pi a H)$. The expression~\eqref{peakhorizon} gives 
$x_p \approx (3w+1)/2$, which yields $x_p \approx 1.5$ and $x_p \approx 2$ for the cases $w=2/3$ and $w=1$ studied there, respectively. It is to be compared with $x_p =1.64 \pm 0.09$ and $x_p =2.16 \pm 0.11$ in Table~5 of Ref.~\cite{Blasi:2025tmn}.} Ref.~\cite{Blasi:2025tmn}.
The expression~\eqref{peakhorizon}  reiterates one of our major points that the particle horizon rather than the inverse Hubble rate is fundamental for DW evolution. 
In the case $w \gg 1$ shown in Fig.~\ref{plateau}, the particle horizon still sets the characteristic wavelength of GWs, which is, however, associated with the origin of a plateau rather than a peak (the latter is virtually absent for $w \gg 1$). 
We comment more on the plateau later and argue for its physical origin. 

Another quantity of interest is the spectral energy density at peak parametrized as 
\begin{equation}
\Omega_{gw, peak} = \frac{\tilde{\epsilon}_{gw} \xi^2 \sigma^2_{wall}}{24 \pi H^2 (\tau) M^4_{P}} \; .
\end{equation}
This expression may serve as the definition of $\tilde{\epsilon}_{gw}$ quantifying efficiency of GW emission~\cite{Hiramatsu:2013qaa}. The quantity $\tilde{\epsilon}_{gw}$ is not universal for different background cosmologies. In particular, one has $\tilde{\epsilon}_{gw} \approx (0.17, 0.25, 0.41, 0.75, 0.79)$ for the EoS parameter $w=(1/3, 1, 5/3, 13/3, 61/3)$. We have taken the value of $\tilde{\epsilon}_{gw}$ in the case of RD stage from Ref.~\cite{Dankovsky:2025pjg}. Hence, efficiency of GWs grows with $w$, but appears to reach saturation eventually. Let us compare it with the theoretical prediction inferred from Eq.~\eqref{theor}. One finds that the values of $\tilde{\epsilon}_{gw}$ are related as $0.11:0.25:0.36:0.60:0.88$ for $w=(1/3, 1, 5/3, 13/3, 61/3)$. 
We observe a qualitatively good agreement, but the theoretical value of $\tilde{\epsilon}_{gw}$ seems to be somewhat underestimated for $w=1/3$.

\subsection{Gravitational wave spectra from annihilating domain walls}
\label{ann}

For the analysis of GWs from DWs annihilating due to the potential bias, we switch to higher resolution lattices with $4096^3$ and $2048^3$ grid points in the case $w=1/3$ and $w >1/3$, respectively. The reason is that annihilating DWs are of considerable phenomenological interest; thus one is motivated to look into the detailed structure of their GW spectra to be able to differentiate them from GWs of the astrophysical origin; see ~\cite{Datta:2026fav, Datta:2026ffs} and references there. In Ref.~\cite{Babichev:2025stm}, we have already performed the analysis in the RD case using $2048^3$ lattice. These results are updated below leveraging not only higher resolution simulations, but also the improved techniques of GW spectra reconstruction discussed below.

There are some caveats on the way to careful reconstruction of GW spectra. 
First, a finite lattice spacing leads to the appearance of an artificial peak in the UV part of the spectrum, whose IR slope affects the central part of the spectrum, see Fig.~\ref{peaksVShubble}. 
Second, there is a considerable uncertainty when estimating IR modes stemming from a finite duration of simulations~\cite{DG}. Namely, in the low $k$ regime such that $k\tau \lesssim 1$ there is no enough time for GWs to undergo a single oscillation. The uncertainty is exacerbated by a limited number of IR modes constrained by $k \geq 2\pi/L$. It is common to ignore this problem and set the IR slope of GWs to be $\Omega_{gw} \propto k^3$ in the RD case. However, such a behavior suggested by causality considerations~\cite{Caprini:2009fx}, is valid only in the limit $k \rightarrow 0$, which is neither accessible by numerical simulations nor by observations. In the realistic case of finite $k$, one naturally gets the departures from the $k^3$-law. Our goal is to capture those departures. 

The strategy to circumvent these technical limitations is as follows:

i) To remove the UV artifact and increase the number of IR modes, we perform simulations using two lattice boxes with different sizes $L_1$ and $L_2$, say $L_1<L_2$. For a larger box with $L_2$, one ameliorates IR properties of the lattice, but impairs its UV properties, and vice versa for a smaller box with $L_1$. We keep only IR modes of the spectrum reconstructed using the $L_2$-box, and only UV modes of the spectrum obtained with the $L_1$-box. Then, we glue the IR and UV modes selected in this way and obtain the upgraded spectrum. The comoving box size is given by Eq.~\eqref{L}, where we again set $\kappa=1$, but choose different $\kappa'=2\pi$ and $\kappa'=\pi/12$ for the larger and the smaller boxes, respectively.

ii) To smoothen the IR part of GW spectrum, we are running simulations for some time after DW network destruction. Such a prolongation of simulations apparently does not cause problems with resolution of DWs, because they disappear at the time $\tau_{ann}$. During the time $\tau_{ann}\lesssim \tau \lesssim \tau_f$, the ``problematic'' IR modes, which are superhorizon by the time of DW annihilation, $k\tau_{ann} \ll 1$, have a chance to undergo at least one oscillation, and we keep only such IR modes. Averaging over the oscillation(s), 
we considerably reduce the uncertainty of determining the IR part of GW spectrum. Details of the procedure are presented in Ref.~\cite{DG}.
We choose the final time of simulations $\tau_f=100$ in the case of RD stage and kination, and $\tau_f=150$ for the universe with $w=5/3$.

\begin{figure}[!htb]
\begin{center}
    \includegraphics[width=\textwidth]{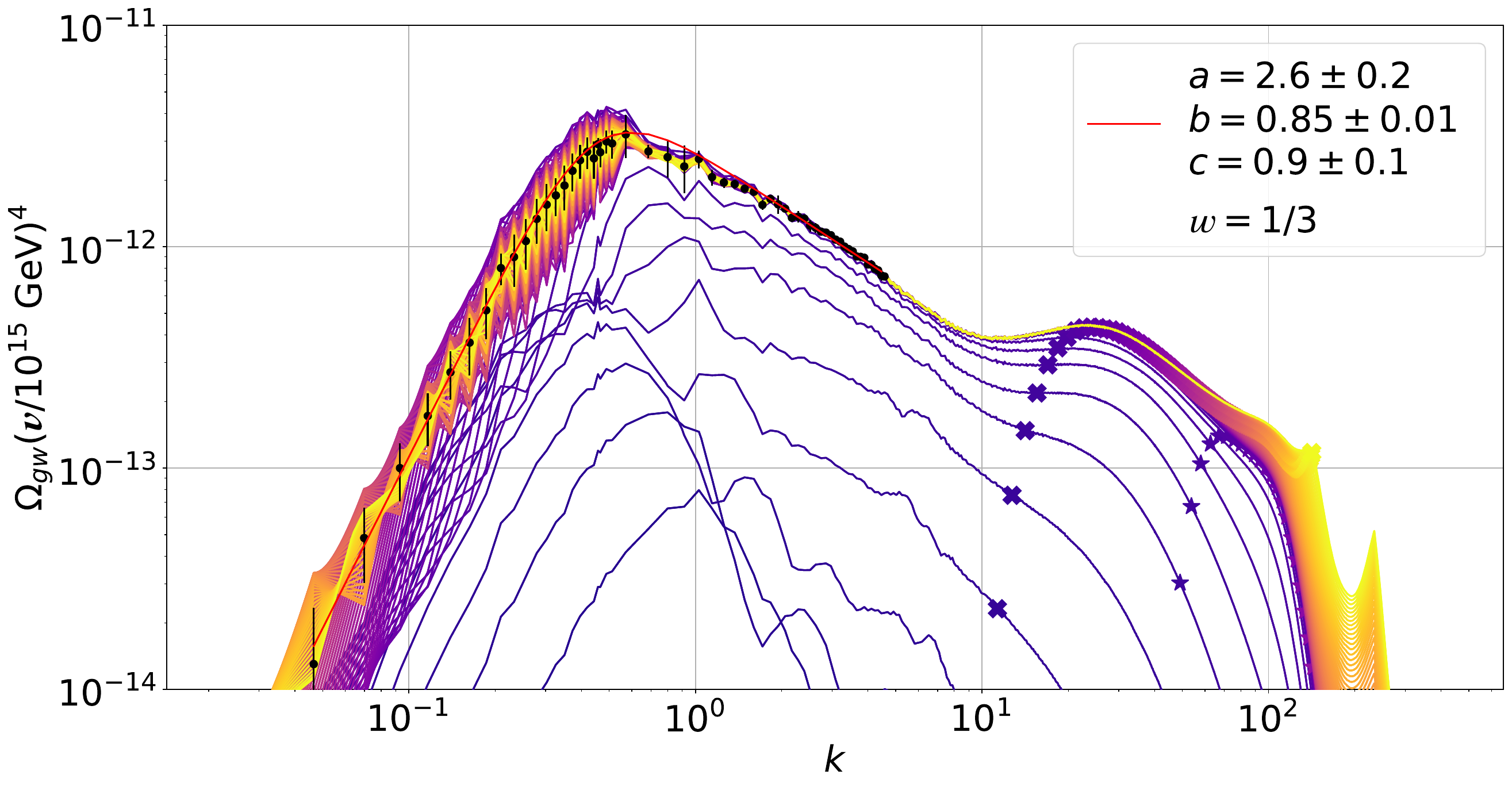} 
\end{center}
    \caption{GW spectra from annihilating DWs reconstructed with $4096^3$ lattice assuming RD universe. The bias parameter is set to $\epsilon=0.025$. The crosses and stars mark the positions corresponding to $k=m_{\chi}a $ and $k =2\pi a/\delta_{wall}$, respectively. Reconstruction has been performed with the use of the procedure discussed in Ref.~\cite{DG}. The fitting is performed using Eq.~\eqref{ansatz}; the goodness-of-fit parameters are given by $\chi_\nu^2 = 1.48$ and ${\cal N}_{dof}=53$.} .\label{GW_spectrum_b_4}
\end{figure}

\begin{figure}[!htb]
\begin{center}
     \includegraphics[width=\textwidth]{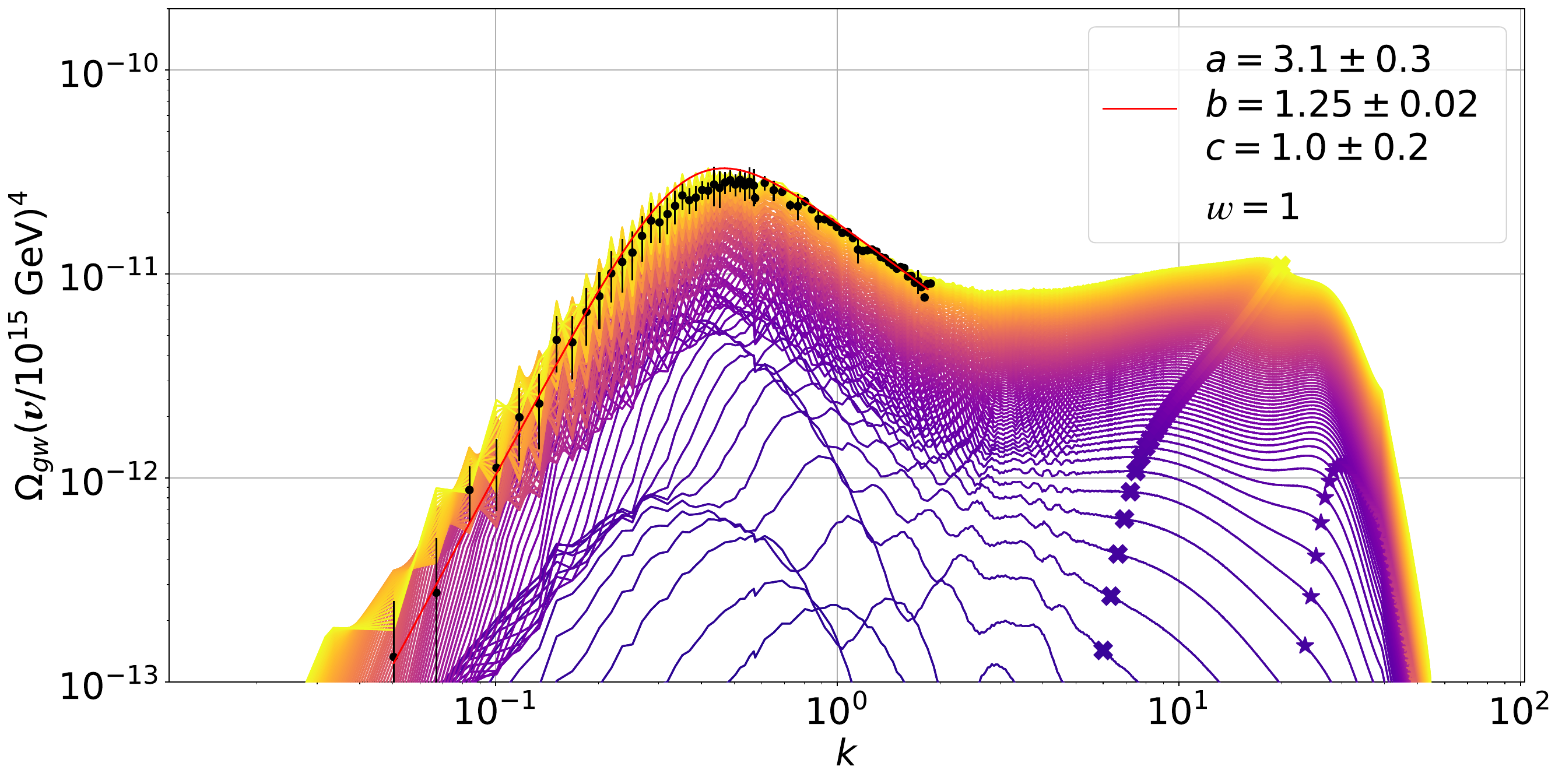} 
\end{center}
    \caption{GW spectra from annihilating DWs reconstructed with $2048^3$ lattice assuming kination, $w=1$. The bias parameter is set to $\epsilon=0.025$. The crosses and stars mark the positions corresponding to $k=m_{\chi}a $ and $k =2\pi a/\delta_{wall}$, respectively. The fitting is performed using Eq.~\eqref{ansatz}; the goodness-of-fit parameters are given by $\chi_\nu^2 = 16.5$ and ${\cal N}_{dof}=62$.}\label{GW_specrtrum_b_6}
\end{figure}

\begin{figure}[!htb]
\begin{center}
       \includegraphics[width=\textwidth]{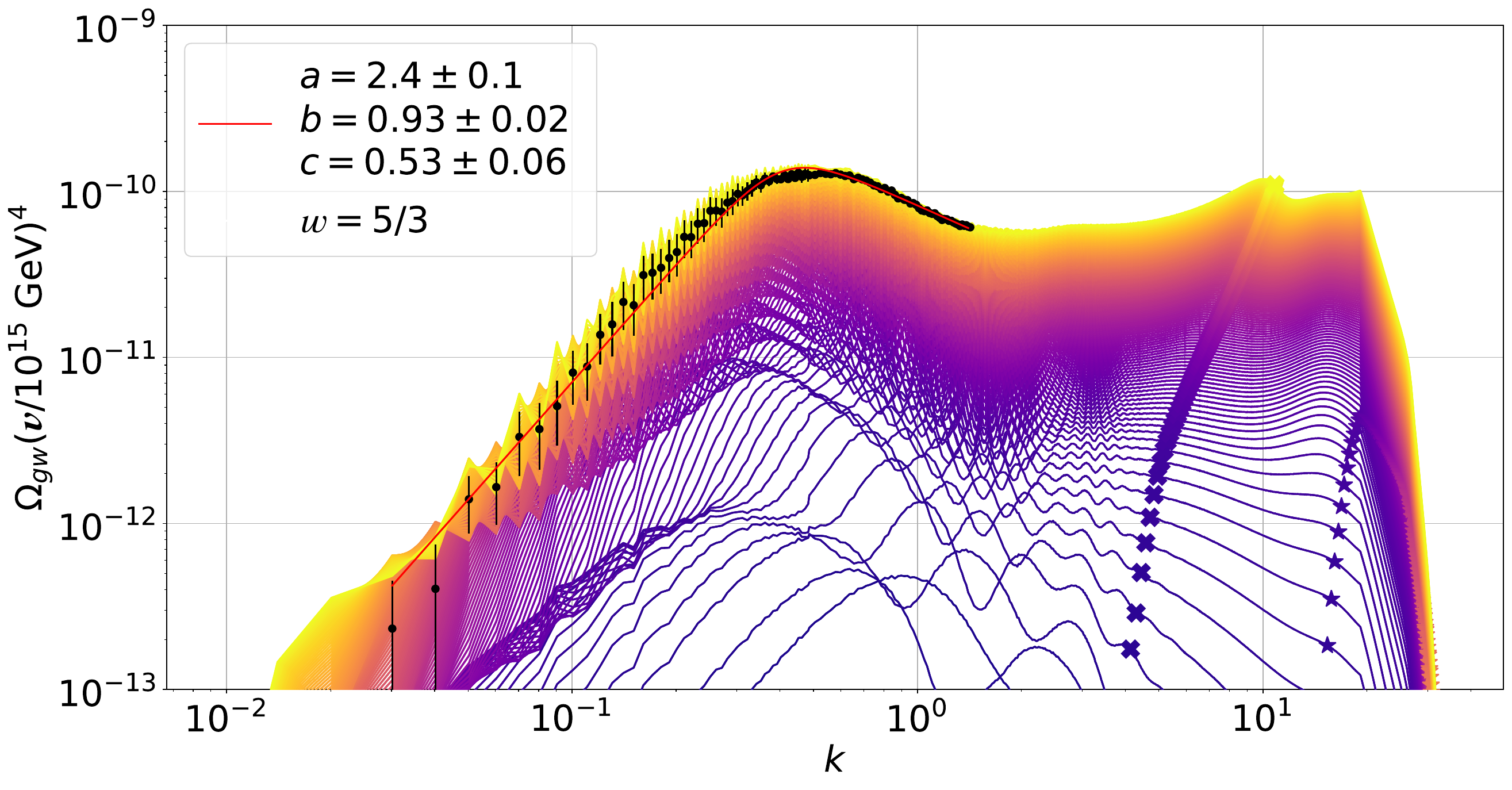} 
\end{center}
    \caption{GW spectra from annihilating DWs reconstructed with $2048^3$ lattice assuming the EoS parameter $w=5/3$. The bias parameter is set to $\epsilon=0.025$. The crosses and stars mark the positions corresponding to $k=m_{\chi}a $ and $k =2\pi a/\delta_{wall}$, respectively. The fitting is performed using Eq.~\eqref{ansatz}; the goodness-of-fit parameters are given by $\chi_{\nu}^2 = 1.24$ and ${\cal N}_{dof}=92$.} 
    \label{GW_spectrum_b_8}
\end{figure}

Results of numerical simulations of GW spectra from annihilating DWs are shown in Figs.~\ref{GW_spectrum_b_4},~\ref{GW_specrtrum_b_6},~\ref{GW_spectrum_b_8}, which correspond to $w=1/3, 1, 5/3$, respectively.  The GW spectrum reaches saturation some time after DW annihilation in the case of RD, which is expected, because the energy density of GWs decreases as radiation, 
$\rho_{gw} \propto 1/a^4$, once the source is switched off. 
On the other hand, $\Omega_{gw}$ continues growing for $w>1/3$, because $\rho_{gw}$ drops slower than $\rho_{U} \propto 1/a^{3(1+w)}$ in that case. 
When fitting GW spectra, we consider the following ansatz:
\begin{equation}
\label{ansatz}
\frac{\Omega_{gw}}{\Omega_{gw, peak}}=\frac{(a+b)^{c}}{\left[b \left(\frac{k_{peak}}{k} \right)^{\frac{a}{c}} +a \left(\frac{k}{k_{peak}} \right)^{\frac{b}{c}} \right]^c} \; ,
\end{equation}
where $a$, $b$, and $c$ are constants. In the literature one typically sets $a=3$, the choice 
supported by the causality enforcing $\Omega_{gw} \propto k^3$ in the limit $k \rightarrow 0$ during RD stage~\cite{Caprini:2009fx}. We should keep the parameter $a$
variable, first because our main focus is on cosmologies different from RD, and second because we are interested in the near peak behavior of the spectrum rather than its deep IR properties.

\begin{figure}[!htb]
\begin{center}
    \includegraphics[width=0.495\textwidth]{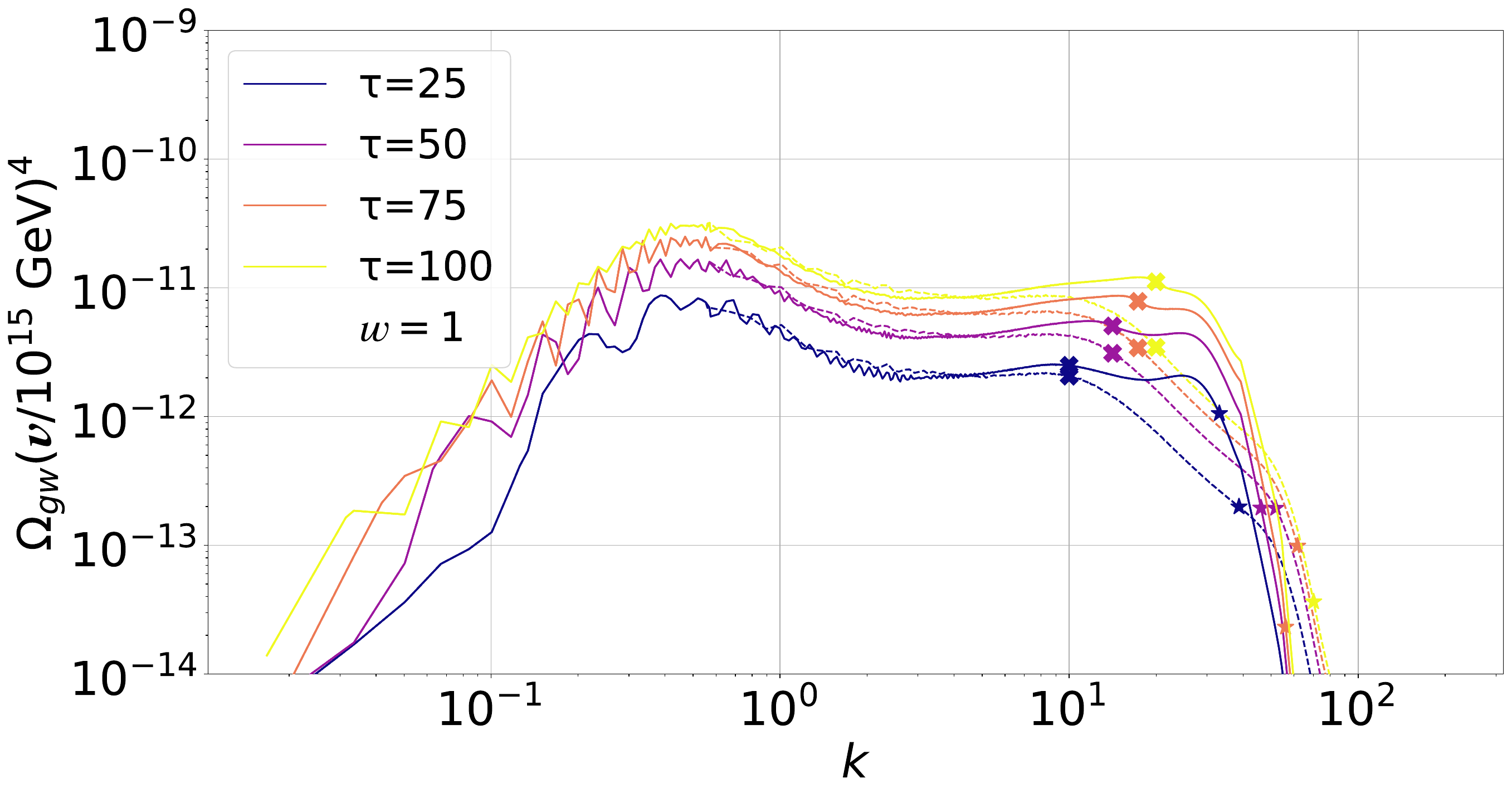}
       \includegraphics[width=0.495\textwidth]{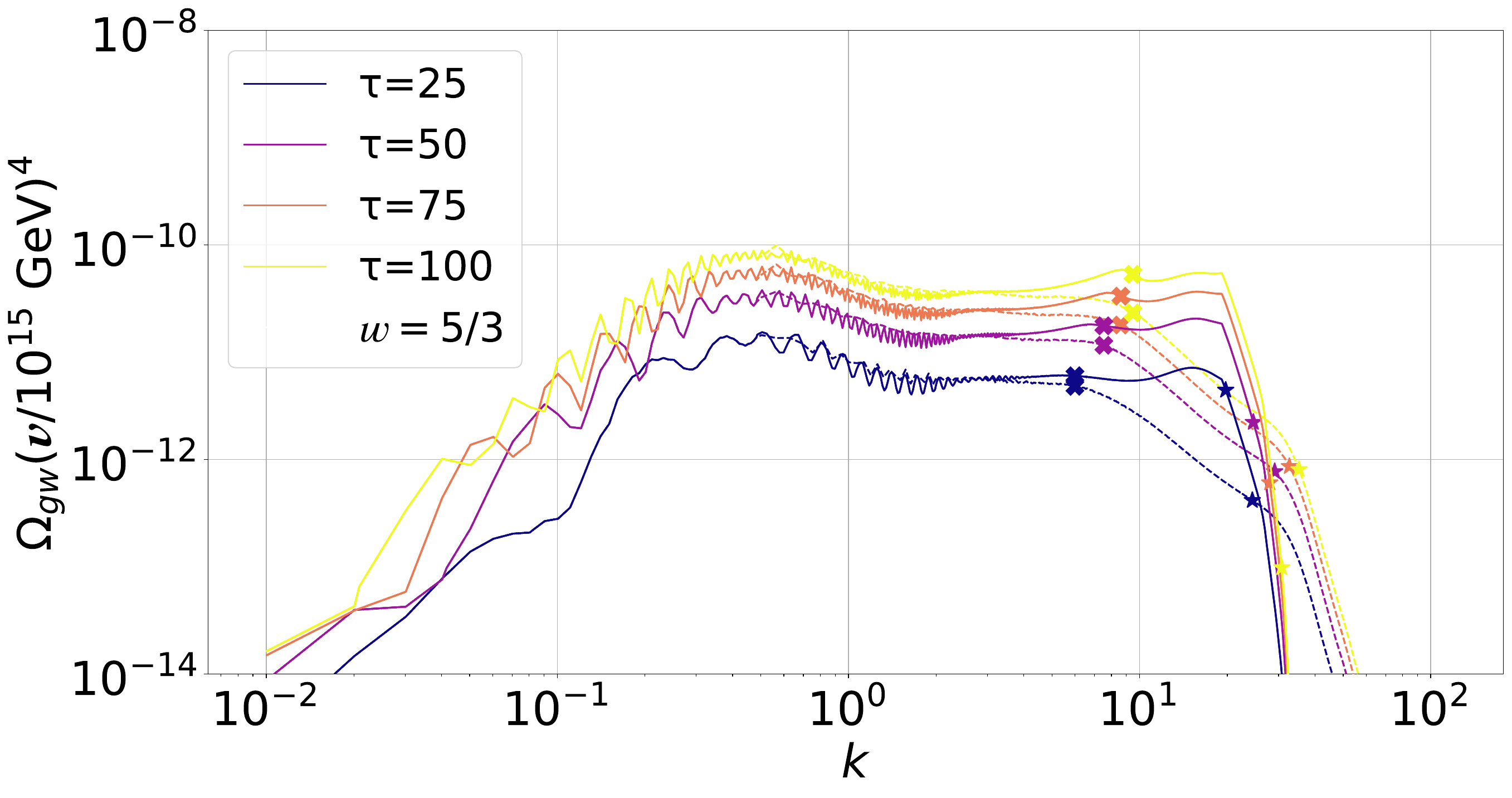} 
\end{center}
    \caption{GW spectra from annihilating DWs described by the bias parameter $\epsilon=0.025$ are shown in the case $w=1$ (left panel) and $w=5/3$ (right panel). Solid lines correspond to GW spectra from Figs.~\ref{GW_specrtrum_b_6} and~\ref{GW_spectrum_b_8} for selected conformal times. Dashed lines correspond to UV parts of GW spectra reconstructed for the same model constants and for the same selected conformal times, but using a lattice box smaller by the factor $2.9$ and $3.3$ in the case $w=1$ and $w=5/3$, respectively. The crosses and stars mark the positions corresponding to $k=m_{\chi} a$ and $k=2\pi a/\delta_{wall}$, respectively.} 
    \label{comparison}
\end{figure}

Below we perform a detailed breakdown of GW spectra obtained.

{\it The IR part}. Here one observes spectral indices different from those given by Eqs.~\eqref{spectrumir} and~\eqref{kinspectrum}. Note that the values of $k$ are bounded in lattice simulations as $k \geq 2\pi/L$, and this is a likely origin of the departure from these theoretical predictions holding in the limit $k \rightarrow 0$. We observe that the fitting parameter $a$ is larger in the RD case, Fig.~\ref{GW_spectrum_b_4}, compared to kination, Fig.~\ref{GW_specrtrum_b_6}, which appears to reflect the deep IR behavior described by Eqs.~\eqref{spectrumir} and~\eqref{kinspectrum}. However, the GW spectrum in Fig.~\ref{GW_spectrum_b_8} does not follow this trend obscuring the connection with the expression $n_{IR}=4-2\nu$.

{\it The near-to-maximum UV part.} To the right of the peak, one observes the power-law decrease of the spectrum $\Omega_{gw} \propto 1/k^{n_{UV}}$, 
where $0.8 \lesssim n_{UV} \lesssim 1.2$. Such a behavior is reminiscent of the power-law decrease $\Omega_{gw} \propto 1/k$ 
observed from bubble collisions during first order phase transitions~\cite{Huber:2008hg}. 
This might be not an accident, because the DW annihilation occurs similarly to the first order phase transition, 
but there is a clear distinction. While during first order phase transitions bubbles of true vacua expand, here one deals 
with pockets of false vacuum shrinking. Therefore, the similarity should not be overestimated. The near-to-maximum UV behavior shown in Fig.~\ref{GW_spectrum_b_4} is in a good agreement with the results obtained in Refs.~\cite{Cyr:2025nzf, Babichev:2025stm}, which studied annihilating DWs in the RD universe. On the other hand, there is a discrepancy with Refs.~\cite{Notari:2025kqq, Barbini:2026edx} exhibiting a broken power law between the peak and the plateau. In Refs.~\cite{Notari:2025kqq, Barbini:2026edx}, the discrepancy is attributed to the scalar field oscillations following the phase transition, when DWs are being formed. These oscillations are suppressed in Refs.~\cite{Notari:2025kqq, Barbini:2026edx} by introducing an artificial friction term (on top of Hubble friction) operating within some early time period. However, this additional friction may not be that innocuous and can be in fact responsible for the observed discrepancy, while the impact of scalar field oscillations on the GW spectrum appears to be somewhat exaggerated. We leave the study of these interesting questions for the future.

{\it Plateau.} The power-law decrease is followed by a plateau already seen in Fig.~\ref{plateau}. This feature has been previously noted in Refs.~\cite{Blasi:2025tmn, Dankovsky:2025pjg, Ferreira:2023jbu, Dankovsky:2024zvs, Kitajima:2023cek} assuming RD stage; still there is no consensus on its nature, i.e., if the plateau has a physical origin or it is caused by some numerical artifact. This question is relevant because of the effects due to the lattice spacing mentioned earlier, which cause contamination in the UV part of the spectrum. We have developed the technique to mitigate such effects, but some contamination may remain and manifest as GW excess. Yet we find it unlikely that the plateau is due to numerical limitations. From the comparison of Figs.~\ref{GW_spectrum_b_4},~\ref{GW_specrtrum_b_6}, and~\ref{GW_spectrum_b_8}, we see that the plateau becomes more pronounced, as one increases the EoS parameter $w$, starting already at $k_{*} \sim (2-3)k_{peak}$ in the particular case $w=5/3$. Such $k_*$ corresponds to the middle of the spectrum, where we do not expect significant resolution issues. Furthermore, in the cases $w=1$ and $w=5/3$ we have performed additional simulations choosing the comoving box size $L'_1$ satisfying $L'_1<L_1<L_2$, see Fig.~\ref{comparison}. The box size $L'_1$ is again given by Eq.~\eqref{L}, where we have set $\kappa=5$ and $\kappa'=\pi/12$. According to the earlier discussion, this allows for a deeper zoom into UV properties of the spectrum. We see in Fig.~\ref{comparison} that only the farthest edge of the plateau is affected\footnote{In this regard, our Fig.~\ref{comparison} agrees well with Fig.~18 of Ref.~\cite{Blasi:2025tmn}. However, compared to Ref.~\cite{Blasi:2025tmn} where the plateau starts at $k \sim 10$, in our case it extends to ``safer'' $k \sim 1$, and we observe that the plateau is unaffected for $1\lesssim k \lesssim 10$. Such an extended plateau is a key advantage achieved by studying the cases with $w>1/3$.}, while its origin and height remain intact. We conclude with the physical nature of the plateau. Owing to the qualitative similarity of Figs.~\ref{GW_spectrum_b_4},~\ref{GW_specrtrum_b_6}, and~\ref{GW_spectrum_b_8}, one extrapolates this conclusion to the 
case $w=1/3$. It is remarkable that increasing $w$ one can sharpen features, which are less pronounced for smaller $w$.
This suggests a promising way of filtering out physical effects by altering $w$. 

As it follows, the plateau is {\it not} confined to the highest GW frequencies of the order of the inverse wall width. Thus, unlike in Ref.~\cite{Barbini:2026edx}, we consider the plateau to be of substantial phenomenological interest. 
It has been suggested in Ref.~\cite{Babichev:2025stm} that the origin of the plateau $k_*$ can be linked to the composite scale $k_* \sim 4\pi a (\tau_{ann})  \sqrt{H(\tau_{ann})/\delta_{wall}}$. In support of this {\it hypothesis} we note that $k_*$ decreases, as one increases $w$ and hence reduces the Hubble rate $H$, --- the tendency, which can be observed in Figs.~\ref{GW_spectrum_b_4},~\ref{GW_specrtrum_b_6}, and~\ref{GW_spectrum_b_8}. The dependence of $k_*$ on $\delta_{wall}$ means violation of the scaling behavior. This appears to be at odds with the analysis of Ref.~\cite{Blasi:2025tmn} demonstrating that DWs exhibit the scaling behavior down to distances of the order of $\delta_{wall}$. This observation of Ref.~\cite{Blasi:2025tmn} has been derived from the numerical study of the equal time correlator defined in Eq.~\eqref{correlator}. On the other hand, the GW spectrum generally receives contributions from the times $\tau_1 \neq \tau_2$. It is plausible that the wall width $\delta_{wall}$ enters the spectrum $P(k, \tau_1, \tau_2)$ through the combination vanishing in the limit $\tau_1 \rightarrow \tau_2$, e.g., $(\tau_1-\tau_2)/\delta_{wall}$. In that case, violation of scaling remains hidden at the level of the equal time correlator, but pops out in the GW spectrum. This may explain the discrepancy from 
the analysis of Ref.~\cite{Blasi:2025tmn}.

Let us speculate on possible physical origins of the plateau. The latter can be sourced by the inhomogeneous distribution of scalar particles produced by the DW network; in fact, such a scalar emission is a prerequisite to maintain DWs in the scaling regime. The energy density of these non-relativistic particles, $\rho_{\chi} \propto 1/a^3$, redshifts faster than the energy density of DWs, $\rho_{wall} \propto 1/(a \tau)$, for $w <1 $, and vice versa for $w >1 $. Kination, $w=1$, stands as a borderline case. This may explain why the plateau tends to "grab" more space on GW plots, as one increases $w$. This explanation suggests that parameters of the plateau and even its existence may depend on the nature of DW particles, i.e., if they are stable or not. Furthermore, note that the energy density $\rho_{\chi}$ depends on the mass of $\chi$-particles $m_{\chi} \sim 1/\delta_{wall}$, i.e., $\rho_{\chi} \propto m_{\chi}$, and this may be the reason for violation of scaling discussed above. It is also worth pointing out the correlation between the height/extent of the plateau and the number of small closed DWs, see Fig.~\ref{closed}. It is not particularly surprising that increasing the number of closed walls by stiffening the cosmological EoS, one should get a stronger GW signal in the UV part of the spectrum. The smallest walls have a characteristic physical radius $r \sim \delta_{wall} \sim 1/m_{\chi}$, which can be related to appearance of the UV peak in Figs.~\ref{GW_specrtrum_b_6} and~\ref{GW_spectrum_b_8} placed at $k_{peak, UV} \approx m_{\chi} a(\tau)$. However, this peak is rather weak and disappears 
upon "cleaning" the UV part of the spectrum, as it follows from Fig.~\ref{comparison}. Therefore, the origin of this "mass" peak remains obscure.

{\it The exponential decrease.} The spectrum ends with a sharp exponential decrease, which takes place at $k \sim 2\pi a/\delta_{wall}$ for $\tau \lesssim \tau_{ann}$ and stays frozen at later times, playing the role of UV cutoff for GW production. 
This is the unambigious physically justified feature of the spectrum, which is also in agreement with other numerical simulations. 

\section{Conclusions}
\label{sec:conclusions}
In this work we have investigated evolution of DW networks assuming different possible cosmic histories. According to Fig.~\ref{scaling}, the DW area parameter is given by $\xi \approx 1.2$ for most cosmologies considered. As it has been discussed in sec.~\ref{sec:scaling}, this approximate universality points to the exceptional role of the particle horizon in evolution of DWs. This is manifest in Fig.~\ref{peaksVShubble} showing that peak frequencies of GW spectra from stable DWs are given by the inverse particle horizon for not very large $w$. It can be possible in principle to distinguish between different cosmologies via observations of GWs emitted by annihilating DWs. There is a simple analytical relation between the properties of the dominant matter in the universe and the deep IR slope of GWs $n_{IR}=4-2\nu$ for $w \neq 1$, where $\nu$ is given by Eq.~\eqref{nu}, cf. Refs.~\cite{Cai:2019cdl, Hook:2020phx}. In the case of kination one should refer to Eq.~\eqref{kinspectrum}. This generalizes the standard $n_{IR}=3$ valid in the RD case~\cite{Caprini:2009fx}. Furthermore, there is a characteristic plateau in the UV part of the spectrum, which becomes more pronounced as one increases the EoS parameter $w$. However, the near peak spectral shape is less sensitive to cosmological expansion. Hence, if the GW signal is weak, so that only the near peak part of the spectrum is observable, it can be challenging to infer information about cosmological expansion from annihilating DWs. On the other hand, for a sufficiently strong GW signal there are better opportunities to determine the EoS parameter due to a broader frequency range accessible.

Let us comment on the possibility of extending our results to a larger scope of cosmologies. Recall that for numerical shortcomings our discussion has been limited to the case $w \geq 0$, but the conclusions are likely to be generic for the universe experiencing deceleration, $w>-1/3$. 
On the other hand, the case of accelerated universe is qualitatively different, as the particle horizon~\eqref{ph} diverges for $w \leq-1/3$. To get an idea on the correlation length in this case, one simply considers replacing the particle horizon by the "DW horizon":
\begin{equation}
l \rightarrow l_{dw} \equiv a \int^{t}_{t_{i}} \frac{dt'}{a(t')} \; ,
\end{equation}
where $t_i$ is the time of DW network formation. The quantity on the r.h.s. here is always finite, and for $t \gg t_i$ it is indistinguishable from the conformal particle horizon in the case of decelerated universe. For the accelerated universe the conformal "DW horizon" is approximately constant and hence $l_{dw} \propto a$. Let us assume that $l_{dw}$ generically defines the DW network correlation length. In that case the DW energy density  $\rho_{wall} \sim \sigma_{wall}/l_{dw}$ 
redshifts as $\rho_{wall} \propto 1/a$ in the accelerated universe. Such a behavior is indeed characteristic for DWs dominating evolution of the universe; this has been proven recently in the VOS framework in Ref.~\cite{Zeng:2026rxt}. 
Furthermore, this behavior is consistent with the fact that DWs are blown away during inflation. 

It is interesting to compare our results of lattice simulations for evolution of DW networks through different cosmologies with analogous results obtained using analytical or semi-analytical methods. In sec.~\ref{sec:scaling}, we have already found a good qualitative agreement with the results obtained in Ref.~\cite{Martins:2016ois} in the VOS framework assuming PRS approach of Ref.~\cite{Press:1989yh}, but there is some discrepancy in the estimates of DW correlation length, which is worth a more detailed study in the future.
Finally, let us note that Ref.~\cite{Hindmarsh:1996xv} adopts the mean field theory approach from the condensed matter physics to make theoretical predictions for the area parameter $\xi$. The results demonstrated in Table I of  Ref.~\cite{Hindmarsh:1996xv} in the case of three spatial dimensions correspond to the area parameter $\xi \approx (0.85, 0.9, 1.0)$ in Minkowski space, RD and MD universe, respectively. These values of $\xi$ are clearly at odds with those shown in Fig.~\ref{1st}. That discrepancy could be partially attributed to possible deficiencies of the estimator~\eqref{estimator}. See Refs.~\cite{Scherrer:1997sq, Li:2025gld} and the comment after Eq.~\eqref{estimator}. However, it still remains challenging to explain the opposite tendencies observed for the area parameter $\xi$ as one increases the EoS parameter $w$ starting from $w=1/3$. In our simulations we observe a slow but steady upward trend for $\xi$, as one approaches Minkowski evolution, whereas Ref.~\cite{Hindmarsh:1996xv} predicts the downward trend. While it would be premature to make decisive conclusions at this stage, it is fulfilling to realize that numerical observations provide useful implications for the development of DW theory. 

\section*{Acknowledgments}
 Numerical simulations have been partly carried out on the cluster Phoebe (CEICO, Prague) and on the cluster of the theoretical division of INR RAS. The largest-grid runs were carried out on the Karolina supercomputer\footnote{\url{https://www.it4i.cz/en/infrastructure/karolina}}, thanks to the support of the Ministry of Education, Youth and Sports of the Czech Republic through the e-INFRA CZ (ID:90254), Pid:FTA-25-47 and Pid:OPEN-37-21. It is a great pleasure to thank Josef Dvo\v ra\v c\'ek for the assistance with numerical simulations. This work was supported in parts by the scientific program of the National Center for Physics and Mathematics, section 5 "Particle Physics and Cosmology", stage 2026-2027. I. D. acknowledges the support by the Foundation for the Advancement of Theoretical Physics and Mathematics ``BASIS''. The work of A. V. was supported by the Czech Science Foundation, GA\v{C}R, project number 24-13079S.

\end{document}